\documentclass[12pt]{article}
\usepackage{a4wide,epsfig,amsmath}

\newcommand{\agt}{\rlap{\lower 3.5 pt \hbox{$\mathchar \sim$}} \raise 1pt
 \hbox {$>$}}
\newcommand{\alt}{\rlap{\lower 3.5 pt \hbox{$\mathchar \sim$}} \raise 1pt
 \hbox {$<$}}

\catcode`@=11
\newcount\@tempcntc
\def\@citex[#1]#2{\if@filesw\immediate\write\@auxout{\string\citation{#2}}\fi
  \@tempcnta\z@\@tempcntb\m@ne\def\@citea{}\@cite{\@for\@citeb:=#2\do
    {\@ifundefined
       {b@\@citeb}{\@citeo\@tempcntb\m@ne\@citea\def\@citea{,}{\bf
?}\@warning
       {Citation `\@citeb' on page \thepage \space undefined}}%
    {\setbox\z@\hbox{\global\@tempcntc0\csname b@\@citeb\endcsname\relax}%
     \ifnum\@tempcntc=\z@ \@citeo\@tempcntb\m@ne
       \@citea\def\@citea{,}\hbox{\csname b@\@citeb\endcsname}%
     \else
      \advance\@tempcntb\@ne
      \ifnum\@tempcntb=\@tempcntc
      \else\advance\@tempcntb\m@ne\@citeo
      \@tempcnta\@tempcntc\@tempcntb\@tempcntc\fi\fi}}\@citeo}{#1}}
\def\@citeo{\ifnum\@tempcnta>\@tempcntb\else\@citea\def\@citea{,}%
  \ifnum\@tempcnta=\@tempcntb\the\@tempcnta\else
   {\advance\@tempcnta\@ne\ifnum\@tempcnta=\@tempcntb \else
\def\@citea{--}\fi
    \advance\@tempcnta\m@ne\the\@tempcnta\@citea\the\@tempcntb}\fi\fi}
\catcode`@=12

\begin{document}

\title{
\vskip-3cm{\baselineskip14pt
\centerline{\normalsize DESY 13--079\hfill ISSN 0418-9833}
\centerline{\normalsize May 2013\hfill}}
\vskip1.5cm
Average gluon and quark jet multiplicities at higher orders}

\author{Paolo Bolzoni,$\!^1$ Bernd A. Kniehl,$\!^1$
Anatoly V. Kotikov$^{\,1,2}$\\
\\
{\normalsize $^1$ II. Institut f\"ur Theoretische Physik, Universit\"at
Hamburg,}\\
{\normalsize Luruper Chaussee 149, 22761 Hamburg, Germany}\\
\\
{\normalsize $^2$ Bogoliubov Laboratory of Theoretical Physics,}\\ 
{\normalsize Joint Institute for Nuclear Research, 141980 Dubna, Russia}
}

\date{}

\maketitle

\begin{abstract}
We develop a new formalism for computing and including both the perturbative
and nonperturbative QCD contributions to the scale evolution of average gluon
and quark jet multiplicities. 
The new method is motivated by recent progress in timelike small-$x$
resummation obtained in the $\overline{\rm MS}$ factorization scheme. 
We obtain next-to-next-to-leading-logarithmic (NNLL) resummed expressions,
which represent generalizations of previous analytic results.
Our expressions depend on two nonperturbative parameters with clear and simple
physical interpretations.
A global fit of these two quantities to all available experimental data sets
that are compatible with regard to the jet algorithms demonstrates by its
goodness how our results solve a longstandig problem of QCD.
We show that the statistical and theoretical uncertainties both do not exceed
5\% for scales above 10~GeV. 
We finally propose to use the jet multiplicity data as a new way to extract the
strong-coupling constant.
Including all the available theoretical input within our approach, we obtain 
$\alpha_s^{(5)}(M_z)=0.1199\pm 0.0026$ in the $\overline{\rm MS}$ scheme in an
approximation equivalent to next-to-next-to-leading order enhanced by the
resummations of $\ln x$ terms through the NNLL level and of $\ln Q^2$ terms by
the renormalization group, in excellent agreement with the present world
average.

\medskip

\end{abstract}

\newpage

\section{Introduction}
\label{sec:intro}

The production of hadrons is due to the strong interactions of quarks and
gluons.
Quantum chromodynamics (QCD), the gauge theory of the strong interactions,
provides a quantitative description of the transitions from quarks and gluons
to jets of hadrons, which may be tested experimentally. 
When jets are produced at colliders, they can be initiated either by a quark or
a gluon.
The two types of jets are expected to exhibit different properties, above all 
because quarks and gluons carry different color charges and spin. 
In fact, a gluon jet is typically broader and contains a larger amount of
hadrons.
Jets with different mother partons can also be studied by looking for the jet
charge distribution as discussed in Ref.~\cite{Waalewijn:2012sv}, with
important consequences for the physics at the CERN Large Hadron Collider (LHC).
To understand the interplay of quarks and gluons in a jet and to predict
testable consequences thereof lies at the very core of QCD.

The typical way to depict the production of a jet from a parton (quark or
gluon) is the following.
An initial parton starts radiating gluons, which in turn can radiate further
gluons or split into secondary quark-antiquark pairs.
This so-called parton showering process causes the virtualities of the parent
partons to decrease.
Finally, when the virtuality falls below a certain cutoff, the cascade stops
and the final-state partons hadronize into color-neutral hadrons, a process
usually described by phenomenological models.
This happens because the production of hadrons is a typical process where  
nonperturbative phenomena are involved.
However, for particular observables, this problem can be avoided.
In particular, the {\it counting} of hadrons in a jet that is initiated at a
certain scale $Q$ belongs to this class of observables.
In this case, one can adopt with quite high accuracy the hypothesis of Local
Parton-Hadron Duality (LPHD), which simply states that parton distributions are
renormalized in the hadronization process without changing their shapes
\cite{Azimov:1984np}.
Hence, if the scale $Q$ is large enough, this would in principle allow
perturbative QCD to be predictive without the need to consider phenomenological
models of hadronization.
Nevertheless, such processes are dominated by soft-gluon emissions, and it is a
well-known fact that, in such kinematic regions of phase space, fixed-order
perturbation theory fails, rendering the usage of resummation techniques
indispensable.
As we shall see, the computation of avarage jet multiplicities indeed requires
small-$x$ resummation, as was already realized a long time ago
\cite{Mueller:1981ex}.  
In Ref.~\cite{Mueller:1981ex}, it was shown that the singularities for
$x\sim 0$, which are encoded in large logarithms of the kind $1/x\ln^k(1/x)$, 
spoil perturbation theory, and also render integral observables in $x$
ill-defined, disappear after resummation.
Usually, resummation includes the singularities from all orders according to a
certain logarithmic accuracy, for which it {\it restores} perturbation theory.

Small-$x$ resummation has recently been carried out for timelike splitting
fuctions in the $\overline{\mathrm{MS}}$ factorization scheme, which is
generally preferable to other schemes, yielding fully analytic expressions.
In a first step, the next-to-leading-logarithmic (NLL) level of accuracy has
been reached \cite{Vogt:2011jv,Albino:2011cm}.
In a second step, this has been pushed to the
next-to-next-to-leading-logarithmic (NNLL), and partially even to the
next-to-next-to-next-to-leading-logarithmic (N$^3$LL), level \cite{Kom:2012hd}.
Thanks to these results, we are able to analytically compute the NNLL
contributions to the evolutions of the average gluon and quark jet
multiplicities with normalization factors evaluated to next-to-leading (NLO)
and approximately to next-to-next-to-next-to-order (N$^3$LO) in the
$\sqrt{\alpha_s}$ expansion.
The previous literature contains a NLL result on the small-$x$ resummation of
timelike splitting fuctions obtained in a massive-gluon scheme.
Unfortunately, this is unsuitable for the combination with available
fixed-order corrections, which are routinely evaluated in the
$\overline{\mathrm{MS}}$ scheme.
A general discussion of the scheme choice and dependence in this context may
be found in Refs.~\cite{Albino:2011bf,Albino:2011si}.

The average gluon and quark jet multiplicities, which we denote as  
$\langle n_h(Q^2)\rangle_{g}$ and $\langle n_h(Q^2)\rangle_{q}$, respectively,
represent the avarage numbers of hadrons in a jet initiated by a gluon or a
quark at scale $Q$.
In the past, analytic predictions were obtained by solving the equations for
the generating functionals in the modified leading-logarithmic approximation
(MLLA) in Ref.~\cite{Capella:1999ms} through N$^3$LO in the expansion
parameter $\sqrt{\alpha_s}$, i.e.\ through $\mathcal{O}(\alpha_s^{3/2})$.
However, the theoretical prediction for the ratio
$r(Q^2)=\langle n_h(Q^2)\rangle_g/\langle n_h(Q^2)\rangle_q$ given in
Ref.~\cite{Capella:1999ms} is about 10\% higher than the experimental data at
the scale of the $Z^0$ boson, and the difference with the data becomes even
larger at lower scales, although the perturbative series seems to converge
very well.
An alternative approach was proposed in Ref.~\cite{Eden:1998ig}, where a
differential equation for the average gluon-to-quark jet multiplicity ratio was
obtained in the MLLA within the framework of the colour-dipole model, and the
constant of integration, which is supposed to encode nonperturbative
contributions, was fitted to experimental data.
A constant offset to the average gluon and quark jet multiplicities was also
introduced in Ref.~\cite{Abreu:1999rs}. 

Recently, we proposed a new formalism \cite{Bolzoni:2012ed,Bolzoni:2012ii} that
solves the problem of the apparent good convergence of the perturbative series
and does not require any ad-hoc offset, once the effects due to the
mixing between quarks and gluons are fully included. 
Our result is a generalization of the result obtained in
Ref.~\cite{Capella:1999ms}.
In our new approach, the nonperturbative informations to the gluon-to-quark jet
multiplicity ratio are encoded in the initial conditions of the evolution
equations. 
Motivated by the excellent agreement of our results with the experimental data
found in Ref.~\cite{Bolzoni:2012ii}, we propose here to also use our approach
to extract the strong-coupling constant $\alpha_s(Q_0^2)$ at some reference
scale $Q_0$ and thus extend our analysis by adding an apropriate fit parameter.

The paper is organized as follows.
In Section~\ref{ffs}, we introduce the equations governing the evolution of the
average gluon and quark jet multiplicities with the scale $Q$ at which the jet
is initiated, develop a formalism to solve them, and improve our results by
resummation.
In Section~\ref{multiplicities}, we explain how we can predict the
average-jet-multiplicity evolutions in our framework adding as much as possible
available information on small-$x$ timelike resummation. 
In Section~\ref{analysis}, we fit our resummed formulae to the available
experimental data exctracting the initial conditions for the evolutions, and
discuss the uncertainties coming from both the statistical analysis of the data
and the missing higher-order terms.
In Section~\ref{coupling}, we inject the strong-coupling constant into our
analysis and extract it.
Finally, in Section~\ref{conclusions}, we summarize our conclusions and present
an outlook.

\section{Fragmentation functions and their evolution}
\label{ffs}

When one considers average multiplicity observables, the basic equation is the
one governing the evolution of the fragmentation functions $D_a(x,\mu^2)$ for
the gluon--quark-singlet system $a=g,s$.
In Mellin space, it reads:  
\begin{equation}
\mu^2\frac{\partial}{\partial \mu^2} \left(\begin{array}{l} D_s(\omega,\mu^2) \\ D_g(\omega,\mu^2)
\end{array}\right)
=\left(\begin{array}{ll} P_{qq}(\omega,a_s) & P_{gq}(\omega,a_s) \\
P_{qg}(\omega,a_s) & P_{gg}(\omega,a_s)\end{array}\right)
\left(\begin{array}{l} D_s(\omega,\mu^2) \\ D_g(\omega,\mu^2)
\end{array}\right),
\label{ap}
\end{equation}
where $P_{ij}(\omega,a_s)$, with $i,j=g,q$, are the timelike splitting
functions, $\omega=N-1$, with $N$ being the standard Mellin moments with
respect to $x$, and $a_s(\mu^2)=\alpha_s(\mu)/(4\pi)$ is the couplant.
The standard definition of the hadron multiplicities in terms of the 
fragmentation functions is given by their integral over $x$, which clearly corresponds to the first Mellin moment, with $\omega=0$
(see, e.g., Ref.~\cite{Ellis:1991qj}):
\begin{equation}
\langle n_h(Q^2)\rangle_{a}\equiv \left[\int_0^1 dx \,x^\omega
D_a(x,Q^2)\right]_{\omega=0}=D_a(\omega=0,Q^2),
\label{multdef2}
\end{equation}
where $a=g,s$ for a gluon and quark jet, respectively.

The timelike splitting functions $P_{ij}(\omega,a_s)$ in Eq.~(\ref{ap}) may be
computed perturbatively in $a_s$,
\begin{equation}
 P_{ij}(\omega,a_s)=
\sum_{k=0}^\infty a_s^{k+1} P_{ij}^{(k)}(\omega).
\end{equation}
The functions $P_{ij}^{(k)}(\omega)$ for $k=0,1,2$ in the
$\overline{\mathrm{MS}}$ scheme may be found in
Refs.~\cite{Gluck:1992zx,Moch:2007tx,Almasy:2011eq} through NNLO and in 
Refs.~\cite{Vogt:2011jv,Albino:2011cm,Kom:2012hd} with small-$x$ resummation
through NNLL accuracy. 
In the remainder of this section, we explain in detail our new approach to
solve Eq.~(\ref{ap}) in order to use its solution in Eq.~(\ref{multdef2}) to
obtain the average gluon and quark jet multiplicities.
To this end, we first discuss how Eq.~(\ref{ap}) can be diagonalized and then
how to implement resummation to improve it, so as to obtain well-defined
quantities at $\omega=0$.

\subsection{Diagonalization}

It is not in general possible to diagonalize Eq.~(\ref{ap}) because the
contributions to the timelike-splitting-function matrix do not commute at
different orders.
The usual approach is then to write a series expansion about the leading-order
(LO) solution, which can in turn be diagonalized.
One thus starts by choosing a basis in which the timelike-splitting-function
matrix is diagonal at LO (see, e.g., Ref.~\cite{Buras:1979yt}),
\begin{equation}
P(\omega,a_s)=
\left(\begin{array}{ll} P_{++}(\omega,a_s) & P_{-+}(\omega,a_s) 
\\ P_{+-}(\omega,a_s) & P_{--}(\omega,a_s)\end{array}\right)
=a_s\left(\begin{array}{ll} P^{(0)}_{++}(\omega) & 0 
\\ 0 & P^{(0)}_{--}(\omega)\end{array}\right)+a_s^2 P^{(1)}(\omega)
+\mathcal{O}(a_s^3),
\label{pmbasis}
\end{equation} 
with eigenvalues $P_{\pm \pm}^{(0)}(\omega)$.
In one important simplification of QCD, namely ${\mathcal N}=4$ super
Yang-Mills theory, this basis is actually more natural than the $(g,s)$ basis
because the diagonal splitting functions $P^{(k)}_{\pm\pm}(\omega)$ may there be
expressed in all orders of perturbation theory as one universal function with
shifted arguments \cite{Kotikov:2002ab}.

It is convenient to represent the change of basis for the fragmentation
functions order by order for $k\geq 0$ as \cite{Buras:1979yt}:
\begin{eqnarray}
D^+(\omega,\mu_0^2) &=& (1-\alpha_{\omega})D_s(\omega,\mu_0^2) 
- \epsilon_\omega D_g(\omega,\mu_0^2),\nonumber\\
D^-(\omega,\mu_0^2)&=& \alpha_{\omega}D_s(\omega,\mu_0^2) + 
\epsilon_\omega D_g(\omega,\mu_0^2). 
\label{changebasisin}
\end{eqnarray}
This implies for the components of the timelike-splitting-function matrix that
\begin{eqnarray}
P^{(k)}_{--}(\omega) &=& \alpha_\omega  P^{(k)}_{qq}(\omega) 
+ \epsilon_\omega P^{(k)}_{qg}(\omega) + 
\beta_\omega  P^{(k)}_{gq}(\omega)
+ (1-\alpha_\omega)  P^{(k)}_{gg}(\omega), \nonumber \\
 P^{(k)}_{-+}(\omega) &=& P^{(k)}_{--}(\omega) - 
\left(P^{(k)}_{qq}(\omega) +
\frac{1-\alpha_\omega}{\epsilon_\omega}  P^{(k)}_{gq}(\omega) \right), \nonumber \\
P^{(k)}_{++}(\omega) &=&  P^{(k)}_{qq}(\omega) +  P^{(k)}_{gg}(\omega) 
- P^{(k)}_{--}(\omega),
\nonumber \\
 P^{(k)}_{+-}(\omega) &=& P^{(k)}_{++}(\omega) - \left(P^{(k)}_{qq}(\omega) -
\frac{\alpha_\omega}{\epsilon_\omega}  P^{(k)}_{gq}(\omega) \right)
= P^{(k)}_{gg}(\omega) - 
\left(P^{(k)}_{--}(\omega) - \frac{\alpha_\omega}{\epsilon_\omega}  
P^{(k)}_{gq}(\omega) \right),\quad
\label{changebasis}
\end{eqnarray}
where
\begin{equation}
\alpha_\omega=\frac{P_{qq}^{(0)}(\omega)-P_{++}^{(0)}(\omega)}
{P_{--}^{(0)}(\omega)-P_{++}^{(0)}(\omega)},\qquad
\epsilon_\omega=\frac{P_{gq}^{(0)}(\omega)}
{P_{--}^{(0)}(\omega)-P_{++}^{(0)}(\omega)},\qquad
\beta_\omega=\frac{P_{qg}^{(0)}(\omega)}{P_{--}^{(0)}(\omega)-P_{++}^{(0)}(\omega)}.
\label{elements}
\end{equation}

Our approach to solve Eq.~(\ref{ap}) differs from the usual one in that we
write the solution expanding about the diagonal part of the all-order
timelike-splitting-function matrix in the plus-minus basis, instead of its LO
contribution. 
For this purpose, we rewrite Eq.~(\ref{pmbasis}) in the following way:
\begin{equation}
P(\omega,a_s)=
\left(\begin{array}{ll} P_{++}(\omega,a_s) & 0 
\\ 0 & P_{--}(\omega,a_s)\end{array}\right)
+a_s^2 \left(\begin{array}{ll} 0 & P^{(1)}_{-+}(\omega) 
\\ P^{(1)}_{+-}(\omega) & 0\end{array}\right)\
+\left(\begin{array}{ll} 0 & \mathcal{O}(a_s^3) 
\\ \mathcal{O}(a_s^3) & 0\end{array}\right).
\label{sfdec}
\end{equation}

In general, the solution to Eq.~(\ref{ap}) in the plus-minus basis can be
formally written as
\begin{equation}
D(\mu^2)=T_{\mu^2}\left\{\exp{\int_{\mu_0^2}^{\mu^2}\frac{d\bar{\mu}^2}{\bar{\mu}^2}P(\bar{\mu}^2)}
\right\}D(\mu_0^2),
\label{gensol}
\end{equation}
where $T_{\mu^2}$ denotes the path ordering with respect to $\mu^2$ and
\begin{equation}
D=\left(\begin{array}{l} D^+ \\ D^-
\end{array}\right).
\end{equation}
As anticipated, we make the following ansatz to expand about the diagonal part
of the timelike-splitting-function matrix in the plus-minus basis: 
\begin{equation}
T_{\mu^2}\left\{\exp{\int_{\mu_0^2}^{\mu^2}\frac{d\bar{\mu}^2}{\bar{\mu}^2}P(\bar{\mu}^2)}
\right\}
=Z^{-1}(\mu^2)\exp\left[\int_{\mu_0^2}^{\mu^2}\frac{d\bar{\mu}^2}{\bar{\mu}^2}P^{D}
(\bar{\mu}^2)\right]Z(\mu_0^2),
\label{ansatz}
\end{equation}
where
\begin{equation}
P^D(\omega)=
\left(\begin{array}{ll} P_{++}(\omega) & 0 
\\ 0 & P_{--}(\omega)\end{array}\right)
\label{diagpart}
\end{equation}
is the diagonal part of Eq.~(\ref{sfdec}) and $Z$ is a matrix in the
plus-minus basis which has a perturbative expansion of the form
\begin{equation}
Z(\mu^2)=1+a_s(\mu^2)Z^{(1)}+\mathcal{O}(a_s^2).
\label{zpertexp}
\end{equation}
In the following, we make use of the renormalization group (RG) equation for
the running of $a_s(\mu^2)$,
\begin{equation}
\mu^2\frac{\partial}{\partial\mu^2}a_s(\mu^2)=\beta(a_s(\mu^2))
=-\beta_0 a_s^2(\mu^2)
-\beta_1 a_s^3(\mu^2)+\mathcal{O}(a_s^4),
\label{running}
\end{equation}
where
\begin{eqnarray}
\beta_0 &=& \frac{11}{3}C_A - 
\frac{4}{3}n_f T_R,\nonumber\\
\beta_1 &=& \frac{34}{3}C_A^2 - \frac{20}{3}C_A n_f T_R - 
4 C_F n_f T_R,
\end{eqnarray}
with $C_A=3$, $C_F=4/3$, and $T_R=1/2$ being colour factors and $n_f$ being the
number of active quark flavours.
Using Eq.~(\ref{running}) to perform a change of integration variable in
Eq.~(\ref{ansatz}), we obtain
\begin{equation}
T_{a_s}\left\{\exp{\int_{a_s(\mu_0^2)}^{a_s(\mu^2)}\frac{d\bar{a}_s}{\beta(\bar{a}_s)}P(\bar{a}_s)
}
\right\}
=Z^{-1}(a_s(\mu^2))\exp\left[\int_{a_s(\mu_0^2)}^{a_s(\mu^2)}
\frac{d\bar{a}_s}{\beta(\bar{a}_s)}P^{D}
(\bar{a}_s)\right]Z(a_s(\mu_0^2)).
\label{ansatz2}
\end{equation}
Substituting then Eq.~(\ref{zpertexp}) into Eq.~(\ref{ansatz2}),
differentiating it with respect to $a_s$, and keeping only the first term in
the $a_s$ expansion, we obtain the following condition for the $Z^{(1)}$ matrix:
\begin{equation}
Z^{(1)}+\left[\frac{P^{(0)D}}{\beta_0},Z^{(1)}\right]=\frac{P^{(1)OD}}{\beta_0},
\end{equation}
where
\begin{equation}
P^{(1)OD}(\omega)=
\left(\begin{array}{ll} 0 & P^{(1)}_{-+}(\omega) 
\\ P^{(1)}_{+-}(\omega) & 0\end{array}\right).
\end{equation}
Solving it, we find:
\begin{equation}
Z_{\pm\pm}^{(1)}(\omega)=0,\qquad
Z_{\pm\mp}^{(1)}(\omega)=\frac{P_{\pm\mp}^{(1)}(\omega)}{\beta_0+P_{\pm\pm}^{(0)}(\omega)
-P_{\mp\mp}^{(0)}(\omega)}.
\label{zmatrix}
\end{equation}

At this point, an important comment is in order.
In the conventional approach to solve Eq.(\ref{ap}), one expands about the
diagonal LO matrix given in Eq.~(\ref{pmbasis}), while here we expand about the
all-order diagonal part of the matrix given in Eq.~(\ref{sfdec}).
The motivation for us to do this arises from the fact that the functional
dependence of $P_{\pm\pm}(\omega,a_s)$ on $a_s$ is different after resummation.

Now reverting the change of basis specified in Eq.~(\ref{changebasisin}), we
find the gluon and quark-singlet fragmentation functions to be given by
\begin{eqnarray}
D_g(\omega,\mu^2)&=&-\frac{\alpha_\omega}{\epsilon_\omega}D^+(\omega,\mu^2)+
\left(\frac{1-\alpha_\omega}{\epsilon_\omega}\right)D^-(\omega,\mu^2),
\nonumber\\
D_s(\omega,\mu^2)&=&D^+(\omega,\mu^2) + D^-(\omega,\mu^2).
\label{inverbasis}
\end{eqnarray}
As expected, this suggests to write the gluon and quark-singlet fragmentation
functions in the following way:
\begin{equation}
D_a(\omega,\mu^2)\equiv D_a^+(\omega,\mu^2)+D_a^-(\omega,\mu^2), \qquad a=g,s,
\label{decomp}
\end{equation} 
where $D_a^+(\omega,\mu^2)$ evolves like a plus component and
$D_a^-(\omega,\mu^2)$ like a minus component.

We now explicitly compute the functions $D_a^\pm(\omega,\mu^2)$ appearing in
Eq.~(\ref{decomp}).
To this end, we first substitute Eq.~(\ref{ansatz}) into Eq.~(\ref{gensol}).
Using Eqs.~(\ref{diagpart}) and (\ref{zmatrix}), we then obtain
\begin{eqnarray}
D^+(\omega,\mu^2)&=&\tilde{D}^+(\omega,\mu_0^2)\hat{T}_+(\omega,\mu^2,\mu_0^2)
-a_s(\mu^2)Z^{(1)}_{-+}(\omega)\tilde{D}^-(\omega,\mu_0^2)\hat{T}_-(\omega,\mu^2,\mu_0^2),
\nonumber\\
D^-(\omega,\mu^2)&=&\tilde{D}^-(\omega,\mu_0^2)\hat{T}_-(\omega,\mu^2,\mu_0^2)
-a_s(\mu^2)Z^{(1)}_{+-}(\omega)\tilde{D}^+(\omega,\mu_0^2)\hat{T}_+(\omega,\mu^2,\mu_0^2),
\label{result1}
\end{eqnarray}
where 
\begin{equation}
\tilde{D}^\pm(\omega,\mu_0^2) = D^\pm(\omega,\mu_0^2)
+a_s(\mu_0^2) Z_{\mp\pm}^{(1)}(\omega) D^\mp(\omega,\mu_0^2) ,
\label{renfact}
\end{equation}
and
\begin{equation}
\hat{T}_{\pm}(\omega,\mu^2,\mu_0^2)  
= \exp \left[\int^{a_s(\mu^2)}_{a_s(\mu_0^2)}
\frac{d\bar{a}_s}{\beta(\bar{a}_s)} \, 
P_{\pm\pm}(\omega,\bar{a}_s) \right] .
\label{rengroupexp}
\end{equation}
has a RG-type exponential form.
Finally, inserting Eq.~(\ref{result1}) into Eq.~(\ref{inverbasis}), we find
by comparison with Eq.~(\ref{decomp}) that
\begin{equation}
D_a^\pm(\omega,\mu^2)=\tilde{D}_a^\pm(\omega,\mu_0^2)
\hat{T}_{\pm}(\omega,\mu^2,\mu_0^2)
\, H_{a}^\pm(\omega,\mu^2),
\label{evolsol}
\end{equation}
where
\begin{eqnarray}
\tilde{D}_g^+(\omega,\mu_0^2)&=&-\frac{\alpha_\omega}{\epsilon_\omega}
\tilde{D}_s^+(\omega,\mu_0^2),\qquad
\tilde{D}_g^-(\omega,\mu_0^2)=\frac{1-\alpha_\omega}{\epsilon_\omega}
\tilde{D}_s^-(\omega,\mu_0^2),\nonumber\\
\tilde{D}_s^+(\omega,\mu_0^2)&=&\tilde{D}^+(\omega,\mu_0^2),\qquad
\tilde{D}_s^-(\omega,\mu_0^2)=\tilde{D}^-(\omega,\mu_0^2),
\label{rlo}
\end{eqnarray}
and $H_a^\pm(\omega,\mu^2)$ are perturbative functions given by
\begin{equation}
H_a^\pm(\omega,\mu^2) =1- a_s(\mu^2)
Z_{\pm\mp,a}^{(1)}(\omega)+\mathcal{O}(a_s^2).
\label{pertfun}
\end{equation}
At $\mathcal{O}(\alpha_s)$, we have
\begin{equation}
Z_{\pm\mp,g}^{(1)}(\omega)=-Z_{\pm\mp}^{(1)}(\omega)
{ \left(\frac{1-\alpha_\omega}{\alpha_\omega}\right)}^{\pm 1},\qquad
Z_{\pm\mp,s}^{(1)}(\omega)=Z_{\pm\mp}^{(1)}(\omega),
\end{equation}
where $Z_{\pm\mp}^{(1)}(\omega)$ is given by Eq.~(\ref{zmatrix}).

\subsection{Resummation}

As already mentioned in Section~\ref{sec:intro}, reliable computations of
average jet multiplicities require resummed analytic expressions for the
splitting functions because one has to evaluate the first Mellin moment
(corresponding to $\omega=N-1=0$), which is a divergent quantity in the
fixed-order perturbative approach.
As is well known, resummation overcomes this problem, as demonstrated in the
pioneering works by Mueller \cite{Mueller:1981ex} and others 
\cite{Ermolaev:1981cm,Dokshitzer:1982xr,Dokshitzer:1982fh,Dokshitzer:1982ia}.

In particular, as we shall see in Section~\ref{multiplicities}, resummed
expressions for the first Mellin moments of the timelike splitting functions
in the plus-minus basis appearing in Eq.~(\ref{pmbasis}) are required in our
approach.
Up to the NNLL level in the $\overline{\mathrm{MS}}$ scheme, these may be
extracted from the available literature
\cite{Mueller:1981ex,Vogt:2011jv,Albino:2011cm,Kom:2012hd} in closed analytic
form using the relations in Eq.~(\ref{changebasis}).
Note that the expressions are generally simpler in the plus-minus basis,%
\footnote[1]{In fact, one can see from Eq.~(3.3) of Ref.~\cite{Kom:2012hd} that
the resummation of the combination $P_{gg}(\omega,a_s)+P_{qq}(\omega,a_s)$, which
according to Eq.~(\ref{changebasisin}) gives $P_{++}(\omega,a_s)$ because
$P_{--}(\omega,a_s)$ does not need resummation, is much simpler than that of
$P_{gg}(\omega,a_s)$ alone.}
while the corresponding results for the resummation of $P_{gg}(\omega,a_s)$ and
$P_{gq}(\omega,a_s)$ can be highly nontrivial and complicated in higher orders
of resummation.
An analogous observation was made for the double-logarithm aymptotics in the
Kirschner-Lipatov approach \cite{Kirschner:1982xw,Kirschner:1983di}, where 
the corresponding amplitudes obey nontrivial equations, whose solutions are
rather complicated special functions.

For future considerations, we remind the reader of an assumpion already
made in Ref.~\cite{Albino:2011cm} according to which the splitting functions
$P^{(k)}_{--}(\omega)$ and $P^{(k)}_{+-}(\omega)$ are supposed to be free of
singularities in the limit $\omega \to 0$.
In fact, this is expected to be true to all orders.
This is certainly true at the LL and NLL levels for
the timelike splitting functions,
as was verified in our previous work \cite{Albino:2011cm}.
This is also true at the NNLL level, as may be explicitly checked by inserting
the results of Ref.~\cite{Kom:2012hd} in Eq.~(\ref{changebasis}). 
Moreover, this is true through NLO in the spacelike case \cite{Kotikov:1998qt}
and holds for the LO and NLO singularities \cite{Fadin:1998py,Kotikov:2000pm}
to all orders in the framework of the Balitski-Fadin-Kuraev-Lipatov (BFKL)
dynamics \cite{Fadin:1975cb,Kuraev:1976ge,Kuraev:1977fs,Balitsky:1978ic},
a fact that was exploited in various approaches (see, e.g., 
Refs.~\cite{Ciafaloni:2007gf,Altarelli:2008aj} and references cited therein).
We also note that the timelike splitting functions share a number of simple 
properties with their spacelike counterparts.
In particular, the LO splitting functions are the same, and the diagonal
splitting functions grow like $\ln \omega$ for $\omega \to \infty$
at all orders.
This suggests the conjecture that the double-logarithm resummation in the
timelike case and the BFKL resummation in the spacelike case are only related
via the plus components. 
The minus components are devoid of singularities as $\omega \to 0$ and thus 
are not resummed.
Now that this is known to be true for the first three orders of resummation,
one has reason to expect this to remain true for all orders.


Using the relationships between the components of the splitting functions in
the two bases given in Eq.~(\ref{changebasis}), we find that the absence of
singularities for $\omega=0$ in $P_{--}(\omega,a_s)$ and $P_{+-}(\omega,a_s)$
implies that the singular terms are related as
\begin{eqnarray}
P_{gq}^{\rm sing}(\omega,a_s)&=&
-\frac{\epsilon_\omega}{\alpha_\omega}P_{g g}^{\rm sing}(\omega,a_s),
\label{tolja3}\\
P_{qg}^{\rm sing}(\omega,a_s)&=&
-\frac{\alpha_\omega}{\epsilon_\omega}P_{qq}^{\rm sing}(\omega,a_s),
\label{tolja3.1}
\end{eqnarray}
where, through the NLL level,
\begin{equation}
-\frac{\alpha_\omega}{\epsilon_\omega}=\frac{C_A}{C_F}\left[1-\frac{\omega}{6}\left(
1+2\frac{n_f T_R}{C_A}
-4\,\frac{C_F n_f T_R}{C_A^2}\right)\right]+\mathcal{O}(\omega^2).
\label{motivation}
\end{equation}
An explicit check of the applicability of the relationships in
Eqs.~(\ref{tolja3}) and (\ref{tolja3.1}) for $P_{ij}(\omega,a_s)$ with
$i,j=g,g$ themselves is performed in the Appendix.
Of course, the relationships in Eqs.~(\ref{tolja3}) and (\ref{tolja3.1}) may be
used to fix the singular terms of the off-diagonal timelike splitting functions
$P_{qg}(\omega,a_s)$ and $P_{gq}(\omega,a_s)$ using known results for the
diagonal timelike splitting functions $P_{qq}(\omega,a_s)$ and
$P_{gg}(\omega,a_s)$.
Since Refs.~\cite{Vogt:2011jv,Almasy:2011eq} became available during the
preparation of Ref.~\cite{Albino:2011cm}, the relations in Eqs.~(\ref{tolja3})
and (\ref{tolja3.1}) provided an important independent check rather than a
prediction.

We take here the opportunity to point out that Eqs.~(\ref{evolsol}) and
(\ref{rlo}) together with Eq.~(\ref{motivation}) support the motivations for
the numerical effective approach that we used in Ref.~\cite{Bolzoni:2012ed} to
study the average gluon-to-quark jet multiplicity ratio. 
In fact, according to the findings of Ref.~\cite{Bolzoni:2012ed}, 
substituting $\omega=\omega_\mathrm{eff}$, where
\begin{equation}
\omega_\mathrm{eff}=2\sqrt{2C_A a_s},
\label{replacement}
\end{equation}
into Eq.~(\ref{motivation}) exactly reproduces the result for the average
gluon-to-quark jet multiplicity ratio $r(Q^2)$ obtained in
Ref.~\cite{Mueller:1983cq}.
In the next section, we shall obtain improved analytic formulae for the 
ratio $r(Q^2)$ and also for the average gluon and quark jet multiplicities.

Here we would also like to note that, at first sight, the substitution
$\omega=\omega_{\rm eff}$ should induce a $Q^2$ dependence in
Eq.~(\ref{elements}), which should contribute to the diagonalization matrix.
This is not the case, however, because to double-logarithmic accuracy the $Q^2$
dependence of $a_s(Q^2)$ can be neglected, so that the factor
$\alpha_\omega/\epsilon_\omega$ does not recieve any $Q^2$ dependence upon the
substitution $\omega=\omega_{\rm eff}$.
This supports the possibility to use this substitution in our analysis and
gives an explanation of the good agreement with other approaches, e.g.\ that of
Ref.~\cite{Mueller:1983cq}.
Nevertheless, this substitution only carries a phenomenological meaning.
It should only be done in the factor $\alpha_\omega/\epsilon_\omega$, but not
in the RG exponents of Eq.~(\ref{rengroupexp}), where it
would lead to a double-counting problem.
In fact, the dangerous terms are already resummed in Eq.~(\ref{rengroupexp}).

In order to be able to obtain the average jet multiplicities, we have to first 
evaluate the first Mellin momoments of the timelike splitting functions in the
plus-minus basis. 
According to Eq.~(\ref{changebasis}) together with the results given in 
Refs.~\cite{Mueller:1981ex,Kom:2012hd}, we have
\begin{equation}
P_{++}^\mathrm{NNLL}(\omega=0)=\gamma_0(1 - K_1 \gamma_0 + K_2  \gamma_0^2),
\label{nllfirst}
\end{equation}
where
\begin{eqnarray}
\gamma_0&=&P_{++}^\mathrm{LL}(\omega=0)=\sqrt{2 C_A a_s},
\label{llgamma0}\\
K_1&=&\frac{1}{12} 
\left[11 +4\frac{n_f T_R}{C_A} \left(1-\frac{2C_F}{C_A}\right)\right],
\\
K_2&=&\frac{1}{288} 
\left[1193-576\zeta_2 -56\frac{n_f T_R}{C_A} 
\left(5+2\frac{C_F}{C_A}\right)\right]
+ 16 \frac{n^2_f T^2_R}{C^2_A}
\left(1+4\frac{C_F}{C_A}-12\frac{C^2_F}{C^2_A}\right),\quad
\end{eqnarray}
and
\begin{equation}
P_{-+}^\mathrm{NNLL}(\omega=0)=-\frac{C_F}{C_A}\,P_{qg}^{NNLL}(\omega=0),
\label{evolsolaa}
\end{equation}
where
\begin{equation}
P_{qg}^\mathrm{NNLL}(\omega=0) = \frac{16}{3} n_f T_R a_s
-\frac{2}{3} n_f T_R
\left[17-4\,\frac{n_f T_R}{C_A} \left(1-\frac{2C_F}{C_A}\right)\right]
{\left(2C_A a_s^3\right)}^{1/2}.
\label{nllsecondA}
\end{equation}
For the $P_{+-}$ component, we obtain
\begin{equation}
P_{+-}^\mathrm{NNLL}(\omega=0)= \mathcal{O}(a_s^2) .
\end{equation}
Finally, as for the $P_{--}$ component, we note that its LO expression produces
a finite, nonvanishing term for $\omega=0$ that is of the same order in $a_s$
as the NLL-resummed results in Eq.~(\ref{nllfirst}), which leads us to use the
following expression for the $P_{--}$ component: 
\begin{equation}
P_{--}^\mathrm{NNLL}(\omega=0)=-\frac{8n_f T_R C_F}{3 C_A}\,a_s
+ \mathcal{O}(a_s^2),
\label{nllsecond}
\end{equation}
at NNLL accuracy.

We can now perform the integration in Eq.~(\ref{rengroupexp}) through the NNLL
level, which yields
\begin{eqnarray}
\hat{T}_{\pm}^\mathrm{NNLL}(0,Q^2,Q^2_0)&=& 
\frac{T_{\pm}^\mathrm{NNLL}(Q^2)}{T_{\pm}^\mathrm{NNLL}(Q^2_0)},
\label{nnllresult}\\
T_{+}^\mathrm{NNLL}(Q^2) &=& \exp\left\{\frac{4C_A}{\beta_0 \gamma^0(Q^2)}
\left[1+\left(b_1-2C_AK_2\right)a_s(Q^2)\right]\right\}
\left(a_s(Q^2)\right)^{d_+},
\\
T_{-}^\mathrm{NNLL}(Q^2)&=&T_{-}^\mathrm{NLL}(Q^2) =
\left(a_s(Q^2)\right)^{d_-},
\label{nnllresultm}
\end{eqnarray}
where
\begin{equation}
b_1=\frac{\beta_1}{\beta_0},\qquad
d_{-} = \frac{8 n_f T_R C_F}{3 C_A \beta_0},\qquad
d_{+} = \frac{2 C_A K_1}{\beta_0}.
\label{anomdim}
\end{equation}

In order to estimate the contribution to an observable of interest from orders
of perturbation theory beyond our calculation, we may shift the argument of the
strong-coupling constant as
\begin{equation}
a_s(Q^2) \to a_s(\xi Q^2).
\label{shift}
\end{equation}
Applying this shift to Eqs.~(\ref{evolsolaa})--(\ref{anomdim}), there is only
one change in the RG exponents, namely
\begin{equation}
T_{+}^\mathrm{NNLL}(Q^2) = \exp\left\{\frac{4C_A}{\beta_0 \gamma^0(\xi Q^2)}
\left[1+\left(b_1-2C_AK_2
-\frac{\beta_0}{2} \ln \xi \right)
a_s(\xi Q^2)\right]\right\}
\left(a_s(\xi Q^2)\right)^{d_+}.
\label{shiftA}
\end{equation}

\section{Multiplicities}
\label{multiplicities}

According to Eqs.~(\ref{rengroupexp}) and (\ref{evolsol}), the $\pm\mp$
components are not involved in the $Q^2$ evolution of average jet
multiplicities, which is performed at $\omega =0$ using the resummed
expressions for the plus and minus components given in Eq.~(\ref{nllfirst})
and (\ref{nllsecond}), respectively.
We are now ready to define the average gluon and quark jet multiplicities in
our formalism, namely
\begin{equation}
\langle n_h(Q^2)\rangle_a\equiv D_a(0,Q^2)= D_a^+(0,Q^2) + D_a^-(0,Q^2),
\label{multdef}
\end{equation}
with $a=g,s$, respectively.

On the other hand, from Eqs.~(\ref{evolsol}) and (\ref{rlo}), it follows that
\begin{eqnarray}
r_+(Q^2)&\equiv&
\frac{D_g^+(0,Q^2)}{D_s^+(0,Q^2)}=
-\lim_{\omega\rightarrow 0}\frac{\alpha_\omega}{\epsilon_\omega}\,
\frac{H^+_g(\omega,Q^2)}{H^+_s(\omega,Q^2)},
\label{evolsola}\\
r_-(Q^2)&\equiv&\frac{D_g^-(0,Q^2)}{D_s^-(0,Q^2)}=\lim_{\omega\rightarrow 0}
\frac{1-\alpha_\omega}{\epsilon_\omega}\,
\frac{H^-_g(\omega,Q^2)}{H^-_s(\omega,Q^2)}.
\label{rmin}
\end{eqnarray}
Using these definitions and again Eq.~(\ref{evolsol}), we may write general
expressions for the average gluon and quark jet multiplicities:
\begin{eqnarray}
\langle n_h(Q^2)\rangle_g&=&\tilde{D}_g^+(0,Q_0^2)\hat{T}_+^\mathrm{res}(0,Q^2,Q_0^2)
H^+_g(0,Q^2)
\nonumber\\
&&{}+\tilde{D}_s^-(0,Q_0^2)r_-(Q^2)\hat{T}_-^\mathrm{res}(0,Q^2,Q_0^2)
H^-_s(0,Q^2),\nonumber\\
\langle n_h(Q^2)\rangle_s&=&\frac{\tilde{D}_g^+(0,Q_0^2)}{r_+(Q^2)}\hat{T}_+^\mathrm{res}(0,Q^2,Q_0^2)
H^+_g(0,Q^2)\nonumber\\
&&{}+\tilde{D}_s^-(0,Q_0^2)\hat{T}_-^\mathrm{res}(0,Q^2,Q_0^2)
H^-_s(0,Q^2).
\end{eqnarray}
At the LO in $a_s$, the coefficients of the RG exponents are given by
\begin{eqnarray}
r_+(Q^2)&=&\frac{C_A}{C_F},\qquad r_-(Q^2)=0,
\nonumber\\
\qquad H^{\pm}_s(0,Q^2)&=&1,\qquad 
\tilde{D}_a^\pm(0,Q_0^2)=D_a^\pm(0,Q_0^2),
\label{lonnll}
\end{eqnarray}
for $a=g,s$.

It would, of course, be desirable to include higher-order corrections in
Eqs.~(\ref{lonnll}).
However, this is highly nontrivial because the general perturbative structures
of the functions $H^{\pm}_a(\omega,\mu^2)$ and $Z_{\pm\mp,a}(\omega,a_s)$, which
would allow us to resum those higher-order corrections, are presently unknown.
Fortunatly, some approximations can be made.
On the one hand, it is well-known that the plus components by themselves
represent the dominant contributions to both the average gluon and quark jet
multiplicities (see, e.g., Ref.~\cite{Schmelling:1994py} for the gluon case and
Ref.~\cite{Dremin:2000ep} for the quark case).
On the other hand, Eq.~(\ref{rmin}) tells us that $D^-_g(0,Q^2)$ is suppressed
with respect to $D^-_s(0,Q^2)$ because $\alpha_\omega\sim 1+\mathcal{O}(\omega)$.
These two observations suggest that keeping $r_-(Q^2)=0$ also beyond LO should
represent a good approximation.
Nevertheless, we shall explain below how to obtain the first nonvanishing
contribution to $r_-(Q^2)$.
Furthermore, we notice that higher-order corrections to $H^{\pm}_a(0,Q^2)$ and 
$\tilde{D}^\pm_a(0,Q_0^2)$ just represent redefinitions of $D^\pm_a(0,Q_0^2)$ by
constant factors apart from running-coupling effects.
Therefore, we assume that these corrections can be neglected.

Note that the resummation of the $\pm\pm$ components was performed similarly
to Eq.~(\ref{rengroupexp}) for the case of parton distribution functions in
Ref.~\cite{Kotikov:1998qt}.
Such resummations are very important because they reduce the $Q^2$ dependences
of the considered results at fixed order in perturbation theory by properly
taking into account terms that are potentially large in the limit
$\omega \to 0$ \cite{Illarionov:2004nw,Cvetic:2009kw}.
We anticipate similar properties in the considered case, too, which is in line
with our approximations.
Some additional support for this may be obtained from $\mathcal{N}=4$ super
Yang-Mills theory, where the diagonalization can be performed exactly in any
order of perturbation theory because the coupling constant and the
corresponding martices for the diagonalization do not depended on $Q^2$.
Consequently, there are no $Z_{\pm\mp,a}^{(k)}(\omega)$ terms, and only
$P_{\pm\pm}^{(k)}(\omega)$ terms contribute to the integrand of the RG exponent.
Looking at the r.h.s.\ of Eqs.~(\ref{renfact}) and (\ref{pertfun}), we indeed
observe that the corrections of $\mathcal{O}(a_s)$ would cancel each other if
the  coupling constant were scale independent.

We now discuss higher-order corrections to $r_+(Q^2)$.
As already mentioned above, we introduced in Ref.~\cite{Bolzoni:2012ed} an
effective approach to perform the resummation of the first Mellin moment of the
plus component of the anomalous dimension.
In that approach, resummation is performed by taking the fixed-order plus
component and substituting $\omega=\omega_\mathrm{eff}$, where
$\omega_\mathrm{eff}$ is given in Eq.~(\ref{replacement}).
We now show that this approach is exact to $\mathcal{O}(\sqrt{a_s})$.
We indeed recover Eq.~(\ref{llgamma0}) by substituting
$\omega=\omega_\mathrm{eff}$ in the leading singular term of the LO splitting
function $P_{++}(\omega,a_s)$,
\begin{equation}
P^\mathrm{LO}_{++}(\omega)=\frac{4C_Aa_s}{\omega}+\mathcal{O}(\omega^0).
\end{equation}
We may then also substitute $\omega=\omega_\mathrm{eff}$ in
Eq.~(\ref{evolsola}) before taking the limit in $\omega=0$.
Using also Eq.~(\ref{motivation}), we thus find
\begin{equation}
r_+(Q^2)=\frac{C_A}{C_F}\left[1-\frac{\sqrt{2a_s(Q^2) C_A}}{3}\left(
1+2\frac{n_f T_R}{C_A}
-4\frac{C_F n_f T_R}{C_A^2}\right)\right]+\mathcal{O}(a_s),
\label{rplusll}
\end{equation}
which coincides with the result obtained by Mueller in
Ref.~\cite{Mueller:1983cq}.
For this reason and because, in Ref.~\cite{Dremin:1999ji}, the average gluon
and quark jet multiplicities evolve with only one RG exponent, we inteprete
the result in Eq.~(5) of Ref.~\cite{Capella:1999ms} as higher-order
corrections to Eq.~(\ref{rplusll}).
Complete analytic expressions for all the coefficients of the expansion through
$\mathcal{O}(a_s^{3/2})$ may be found in Appendix~1 of
Ref.~\cite{Capella:1999ms}.
This interpretation is also explicitely confirmed in Chapter 7 of
Ref.~\cite{Dokshitzer:1991wu} through $\mathcal{O}(a_s)$.

Since we showed that our approach reproduces exact analytic results at
$\mathcal{O}(\sqrt{a_s})$, we may safely apply it to predict the first
non-vanishing correction to $r_-(Q^2)$ defined in Eq.~(\ref{rmin}), which
yields
\begin{equation}
r_-(Q^2)=-\frac{4 n_f T_R}{3}\sqrt{\frac{2 a_s(Q^2)}{C_A}} +\mathcal{O}(a_s).
\label{frminus}
\end{equation} 
However, contributions beyond $\mathcal{O}(\sqrt{\alpha_s})$ obtained in this
way cannot be trusted, and further investigation is required.
Therefore, we refrain from considering such contributions here.

For the reader's convenience, we list here expressions with numerical
coefficients for $r_+(Q^2)$ through $\mathcal{O}(a_s^{3/2})$ and for $r_-(Q^2)$
through $\mathcal{O}(\sqrt{a_s})$ in QCD with $n_f=5$:
\begin{eqnarray}
r_{+}(Q^2)&=&2.25-2.18249\,\sqrt{a_s(Q^2)}
-27.54\,a_s(Q^2)
+10.8462\,a_s^{3/2}(Q^2)+\mathcal{O}(a_s^2),\label{dreminscaleplus}\\
r_{-}(Q^2)&=&-2.72166\,\sqrt{a_s(Q^2)}+\mathcal{O}(a_s).
\label{dreminscaleminus}
\end{eqnarray}

We denote the approximation in which
Eqs.~(\ref{nnllresult})--(\ref{nnllresultm}) and (\ref{lonnll}) are used as
$\mathrm{LO}+\mathrm{NNLL}$, the improved approximation in which the
expression for $r_+(Q^2)$ in Eq.~(\ref{lonnll}) is replaced by
Eq.~(\ref{dreminscaleplus}), i.e.\ Eq.~(5) in Ref.~\cite{Capella:1999ms}, as
$\mathrm{N}^3\mathrm{LO}_\mathrm{approx}+\mathrm{NNLL}$, and our best
approximation in which, on top of that, the expression for $r_-(Q^2)$ in
Eq.~(\ref{lonnll}) is replaced by Eq.~(\ref{dreminscaleminus}) as
$\mathrm{N}^3\mathrm{LO}_\mathrm{approx}+\mathrm{NLO}+\mathrm{NNLL}$. 
We shall see in Section~\ref{analysis}, where we compare with the
experimental data and extract the strong-coupling constant, that the latter
two approximations are actually very good and that the last one yields the
best results, as expected.

In all the approximations considered here, we may summarize our main
theoretical results for the avarage gluon and quark jet multiplicities in the
following way:
\begin{eqnarray}
\langle n_h(Q^2)\rangle_g&=&n_1(Q_0^2)\hat{T}_+^\mathrm{res}(0,Q^2,Q_0^2)
+n_2(Q_0^2)\,r_-(Q^2)\hat{T}_-^\mathrm{res}(0,Q^2,Q_0^2),
\nonumber\\
\langle n_h(Q^2)\rangle_s&=&n_1(Q_0^2)\frac{\hat{T}_+^\mathrm{res}(0,Q^2,Q_0^2)}
{r_+(Q^2)}+n_2(Q_0^2)\,\hat{T}_-^\mathrm{res}(0,Q^2,Q_0^2),
\label{quarkgen}
\end{eqnarray}
where 
\begin{eqnarray}
n_1(Q_0^2)&=&r_+(Q_0^2)
\frac{D_g(0,Q_0^2)-r_-(Q_0^2)D_s(0,Q_0^2)}{r_+(Q_0^2)-r_-(Q_0^2)},
\nonumber\\
n_2(Q_0^2)&=&\frac{r_+(Q_0^2)D_s(0,Q_0^2)-D_g(0,Q_0^2)}{r_+(Q_0^2)-r_-(Q_0^2)}.
\label{n2}
\end{eqnarray}
The average gluon-to-quark jet multiplicity ratio may thus be written as
\begin{equation}
r(Q^2)\equiv\frac{\langle n_h(Q^2)\rangle_g}{\langle n_h(Q^2)\rangle_s}
=r_+(Q^2)\left[\frac{1+r_-(Q^2)R(Q_0^2)
\hat{T}_{-}^\mathrm{res}(0,Q^2,Q^2_0)/\hat{T}_{+}^\mathrm{res}(0,Q^2,Q^2_0)}
{1+r_+(Q^2)R(Q_0^2)\hat{T}_{-}^\mathrm{res}(0,Q^2,Q^2_0)/
\hat{T}_{+}^\mathrm{res}(0,Q^2,Q^2_0)}\right],
\label{ratiogen}
\end{equation}
where 
\begin{equation}
R(Q_0^2)=\frac{n_2(Q_0^2)}{n_1(Q_0^2)}.
\end{equation}
It follows from the definition of $\hat{T}^\mathrm{res}_\pm(0,Q^2,Q_0^2$ in
Eq.~(\ref{nnllresult}) and from Eq.~(\ref{n2}) that, for $Q^2=Q_0^2$,
Eqs.~(\ref{quarkgen}) and (\ref{ratiogen}) become
\begin{equation}
\langle n_h(Q_0^2)\rangle_g=D_g(0,Q_0^2),\qquad
\langle n_h(Q_0^2)\rangle_q=D_s(0,Q_0^2),\qquad
r(Q_0^2)=\frac{D_g(0,Q_0^2)}{D_s(0,Q_0^2)}.
\label{incond}
\end{equation}
These represent the initial conditions for the $Q^2$ evolution at an arbitrary
initial scale $Q_0$.
In fact, Eq.~(\ref{quarkgen}) is independ of $Q_0^2$, as
may be observed by noticing that
\begin{equation}
\hat{T}_\pm^\mathrm{res}(0,Q^2,Q_0^2)=\hat{T}_\pm^\mathrm{res}(0,Q^2,Q_1^2)
\hat{T}_\pm^\mathrm{res}(0,Q_1^2,Q_0^2),
\end{equation}
for an arbitrary scale $Q_1$ (see also Ref.~\cite{Bolzoni:2012cv} for a
detailed discussion of this point).

In the approximations with $r_-(Q^2)=0$ \cite{Bolzoni:2012ii}, i.e.\ the
$\mathrm{LO}+\mathrm{NNLL}$ and
$\mathrm{N}^3\mathrm{LO}_\mathrm{approx}+\mathrm{NNLL}$ ones, our general results
in Eqs.~(\ref{quarkgen}), and (\ref{ratiogen}) collapse to
\begin{eqnarray}
\langle n_h(Q^2)\rangle_g&=&D_g(0,Q_0^2)\hat{T}_+^\mathrm{res}(0,Q^2,Q_0^2),
\nonumber\\
\langle n_h(Q^2)\rangle_s&=&D_g(0,Q_0^2)
\frac{\hat{T}_+^\mathrm{res}(0,Q^2,Q_0^2)}{r_+(Q^2)}
+\left[D_s(0,Q_0^2)-\frac{D_g(0,Q_0^2)}{{r_+(Q_0^2)}}\right]
\hat{T}_-^\mathrm{res}(0,Q^2,Q_0^2),
\nonumber\\
r(Q^2)  &=& 
\frac{r_{+}(Q^2)}{\left[1 + \frac{r_{+}(Q^2)}{r_{+}(Q_0^2)}\left(
\frac{D_s(0,Q^2_0)r_{+}(Q_0^2)}{D_g(0,Q^2_0)}
-1 \right)
\frac{\hat{T}_{-}^\mathrm{res}(0,Q^2,Q^2_0)}{\hat{T}_{+}^\mathrm{res}(0,Q^2,Q^2_0)}\right]} .
\end{eqnarray}

The NNLL-resummed expressions for the average gluon and quark jet multiplicites
given by Eq.~(\ref{quarkgen}) only depend on two nonperturbative constants,
namely $D_g(0,Q^2_0)$ and $D_s(0,Q^2_0)$.
These allow for a simple physical interpretation.
In fact, according to Eq.~(\ref{incond}), they are the average gluon and quark
jet multiplicities at the arbitrary scale $Q_0$. 
We should also mention that identifying the quantity $r_+(Q^2)$ with the one
computed in Ref.~\cite{Capella:1999ms}, we assume the scheme dependence to be
negligible.
This should be justified because of the scheme independence through NLL  
established in Ref.~\cite{Albino:2011cm}.

We note that the $Q^2$ dependence of our results is always generated via
$a_s(Q^2)$ according to Eq.~(\ref{running}).
This allows us to express Eq.~(\ref{nnllresult}) entirely in terms of
$\alpha_s(Q^2)$.
In fact, substituting the QCD values for the color factors and choosing
$n_f=5$ in the formulae given in Ref.~\cite{Bolzoni:2012ii}, we may write at
NNLL
\begin{eqnarray}
\hat{T}_-^\mathrm{res}(Q^2,Q_0^2)&=&
\left[\frac{\alpha_s(Q^2)}{\alpha_s(Q_0^2)}\right]^{d_1},
\nonumber\\
\hat{T}_+^\mathrm{res}(Q^2,Q_0^2)&=&
\exp\left[d_2\left(\frac{1}{\sqrt{\alpha_s(Q^2)}}
-\frac{1}{\sqrt{\alpha_s(Q_0^2)}}\right)
+d_3\left(\sqrt{\alpha_s(Q^2)}-\sqrt{\alpha_s(Q_0^2)}\right)\right]
\nonumber\\
&&{}\times\left[\frac{\alpha_s(Q^2)}{\alpha_s(Q_0^2)}\right]^{d_4},
\end{eqnarray}
where
\begin{equation}
d_1=0.38647,\qquad
d_2=2.65187,\qquad
d_3=-3.87674,\qquad
d_4=0.97771. 
\end{equation}

We conclude this section by discussing the theoretical uncertainties in
$r_+(Q^2)$ and $r_-(Q^2)$ due to unknown higher-order corrections.
Similarly to Eq.~(\ref{shiftA}), we may estimate them by studying the scale
dependence.
Performing the shift of Eq.~(\ref{shift}) in Eqs.~(\ref{dreminscaleplus}) and
(\ref{dreminscaleminus}), we obtain
\begin{eqnarray}
r_{+}(Q^2)&=&2.25-2.18249\,\sqrt{a_s(\xi Q^2)}
-27.54\,\alpha_s(\xi Q^2)\nonumber\\
&&{}+\left(10.8462-2.18249\,\frac{\beta_0}{2}\ln \xi\right)a_s^{3/2}(\xi Q^2)
+\mathcal{O}(a_s^2),
\nonumber\\
r_{-}(Q^2)&=&-2.72166\,\sqrt{a_s(\xi Q^2)}+\mathcal{O}(a_s).
\label{dreminscale}
\end{eqnarray}

\section{Analysis}
\label{analysis}

We are now in a position to perform a global fit to the available experimental
data of our formulas in Eq.~(\ref{quarkgen}) in the
$\mathrm{LO}+\mathrm{NNLL}$,
$\mathrm{N}^3\mathrm{LO}_\mathrm{approx}+\mathrm{NNLL}$, and 
$\mathrm{N}^3\mathrm{LO}_\mathrm{approx}+\mathrm{NLO}+\mathrm{NNLL}$
approximations, so as to extract the nonperturbative constants $D_g(0,Q^2_0)$
and $D_s(0,Q^2_0)$. 

\begin{figure}
\centering
\includegraphics[width=0.85\textwidth,bb=4 4 313 203]{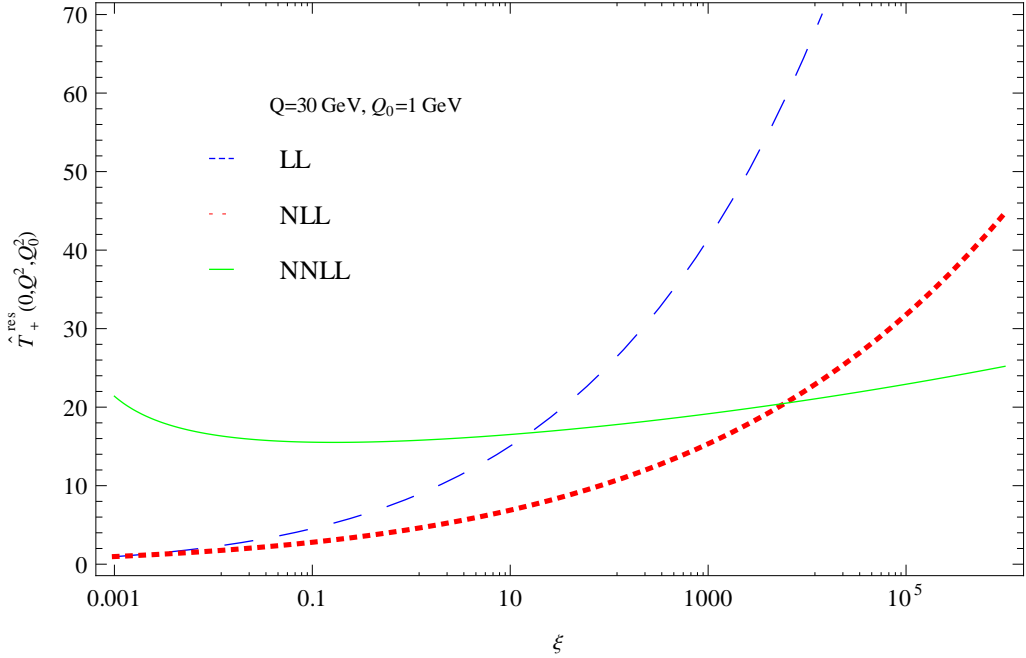}
\caption{\footnotesize{Scale dependences of $\hat{T}_+(0,Q^2,Q_0^2)$ for
$Q=30$~GeV and $Q_0=1$~GeV at the LL (dashed/blue line), NLL (dotted/red line),
and NNLL (continuous/green line) levels.}}
\label{Fig:plotplus1}
\end{figure}

\begin{figure}
\centering
\includegraphics[width=0.85\textwidth,bb=4 4 309 203]{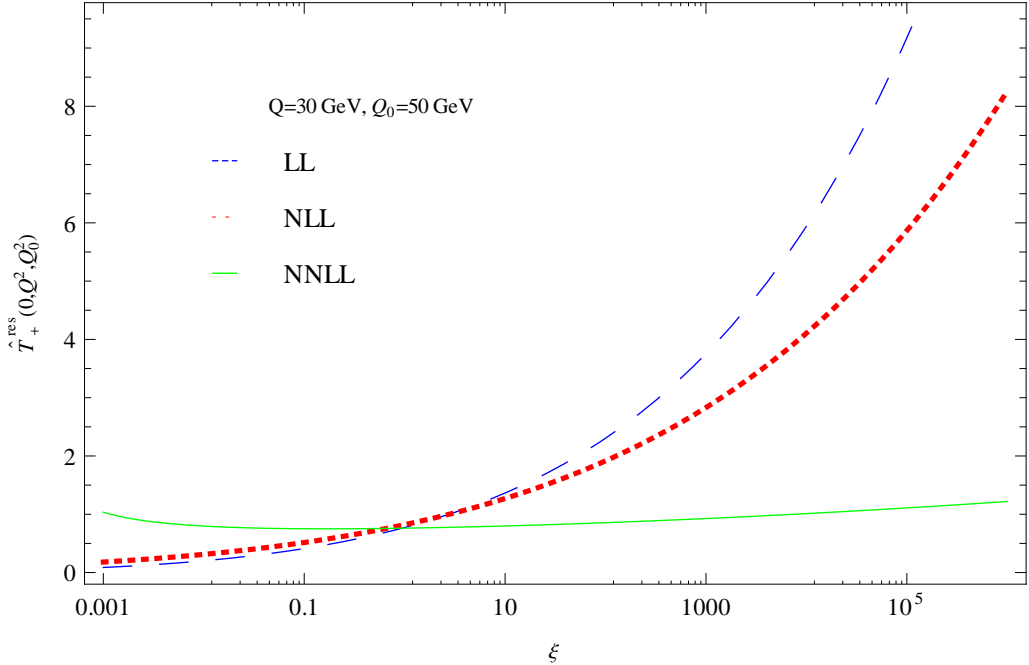}
\caption{\footnotesize{Scale dependences of $\hat{T}_+(0,Q^2,Q_0^2)$ for
$Q=30$~GeV and $Q_0=50$~GeV at the LL (dashed/blue line), NLL (dotted/red line),
and NNLL (continuous/green line) levels.}}
\label{Fig:plotplus2}
\end{figure}

We have to make a choice for the scale $Q_0$, which, in principle, is
arbitrary.
We wish to choose it by optimizing the apparent convergence properties of the
perturbative QCD expansion.
To this end, we analyse in Figs.~\ref{Fig:plotplus1} and \ref{Fig:plotplus2}
the dependence on the scaling parameter $\xi$ of $\hat{T}_+(0,Q^2,Q_0^2)$
governed by Eq.~(\ref{shiftA}) at different logarithmic accuracies for
$Q_0=1$~GeV and $Q_0=50$~GeV, respectively.
We put $Q=30$~GeV because this is in the center of the range where the
majority of the available data located.
We observe a strong reduction of the scale dependence as we pass from LL via
NLL to NNLL, both for $Q_0=1$~GeV and $Q_0=50$~GeV.
The perturbative series appears to be more rapidly converging at relatively
large values of $Q_0$.
Therefore, we adopt $Q_0=50$~GeV in the following.
Another good reason for this choice is that, according to Eq.~(\ref{incond}),
$D_g(0,Q^2_0)$ and $D_s(0,Q^2_0)$ represent the avarage gluon and quark jet
multiplicities, respectively, at the scale $Q_0$, so that the fit results for
our initial conditions may be directly compared with the experimental data
at $Q_0=50$~GeV.

\begin{figure}
\centering
\includegraphics[width=0.85\textwidth,bb=0 1 288 189]{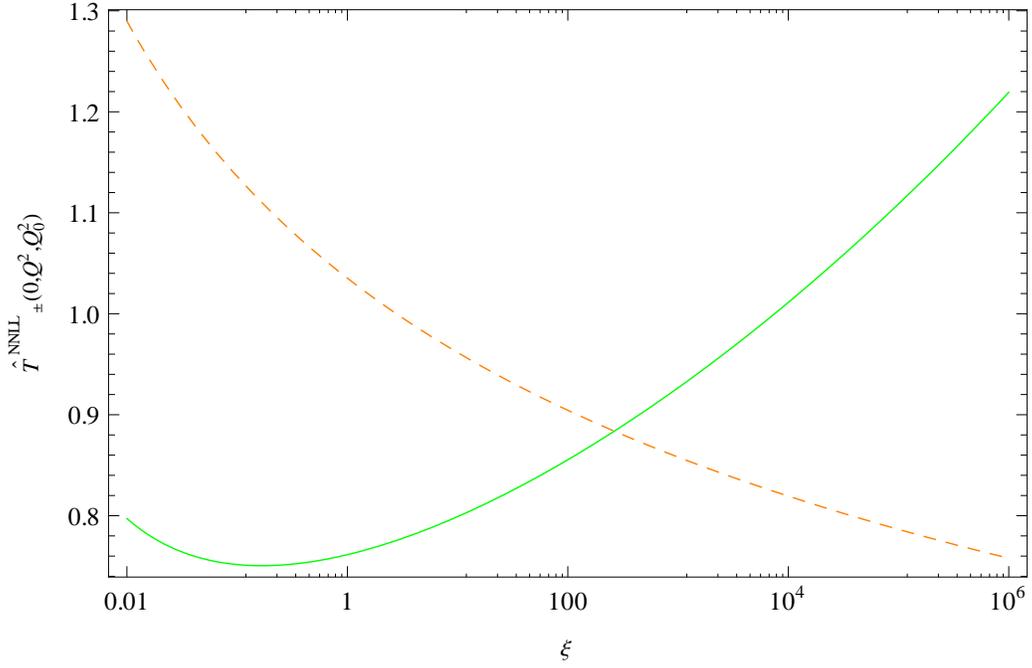}
\caption{\footnotesize{Scale dependences of $\hat{T}_+(0,Q^2,Q_0^2)$
(continuous/green line) and $\hat{T}_-(0,Q^2,Q_0^2)$ (dashed/orange line) for
$Q=30$~GeV and $Q_0=50$~GeV at the NNLL level.}}
\label{Fig:plotminus}
\end{figure}

In Fig.~\ref{Fig:plotminus}, we compare the scale dependence of
$\hat{T}_-(0,Q^2,Q_0^2)$, which is obtained by simply replacing $Q^2$ with
$\xi Q^2$ in Eq.~(\ref{nnllresult}), with the one of
$\hat{T}_+(0,Q^2,Q_0^2)$ evaluated according to Eq.~(\ref{shiftA}), for
$Q=30$~GeV and $Q_0=50$~GeV.
We observe from Fig.~\ref{Fig:plotminus} that the scale variation is very
similar in both cases.

\begin{figure}
\centering
\includegraphics[width=0.85\textwidth,bb=6 9 428 277]{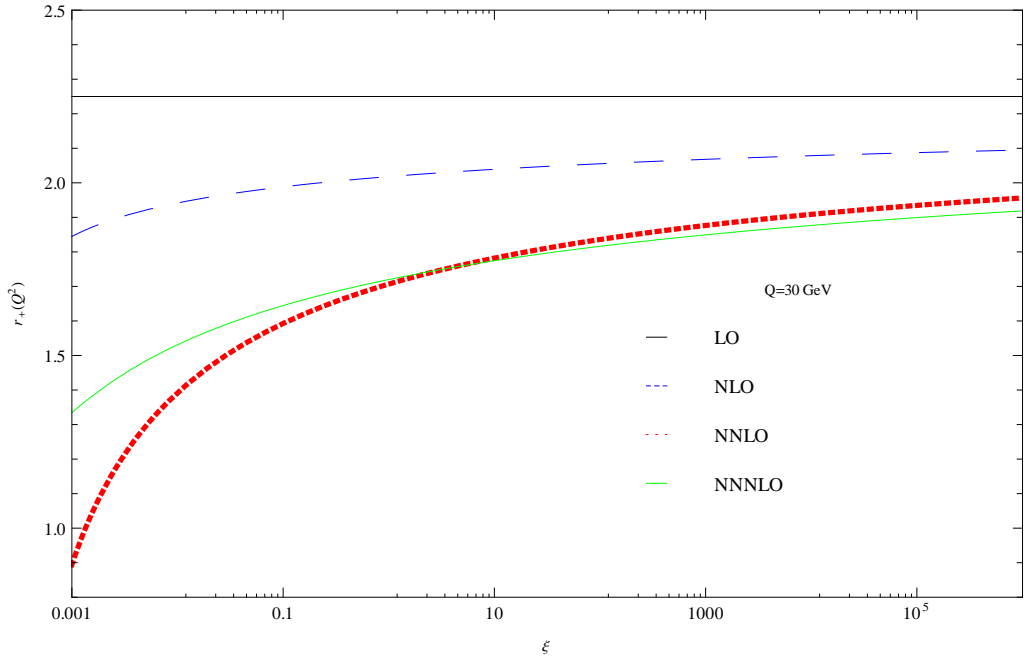}
\caption{\footnotesize{Scale dependence of $r_+(Q^2)$ at LO (dashed/blue line),
NLO (dotted/red line), NNLO (continuous/green line), and N$^3$LO
(continuous/black line).}}
\label{Fig:plotrplus}
\end{figure}

In Fig.~\ref{Fig:plotrplus}, we study the scale dependence of $r_+(Q^2)$
evaluated at LO, NLO, NNLO, and N$^3$LO according to Eq.~(\ref{dreminscale}).
We observe that the scale dependence gradually increases as we pass from LO via
NLO to NNLO, while it decreases in the step from NNLO to N$^3$LO, and hence
conclude that only the latter order may be trusted.

Prior to presenting our fits, we explain our definition of confidence level
(CL), which we adopt from Ref.~\cite{Stump:2001gu}.
Suppose a fit of the free parameters to $n$ experimental data points yields
the minimum $\chi^2$ value $\chi_0^2$.
We then determine the 90\% CL limits on a fit parameter by varying it so that
the resulting $\chi^2$ values stay within the range
\begin{equation}
\chi^2 < \chi_0^2\frac{\zeta_{90}}{\zeta_{50}},
\end{equation} 
where $\zeta_{50(90)}$ are defined such that
\begin{equation}
\int_0^{\zeta_{50(90)}} P(\chi^2,n)\,d\chi^2=0.50(0.90),
\end{equation}
with  
\begin{equation}
P(\chi^2,n)=\frac{2^{-n/2}}{\Gamma(n/2)}\, \left(\chi^2\right)^{n/2-1} e^{-\chi^2/2}.
\end{equation}

\begin{table}
\centering
\begin{tabular}{|c|c|c|c|}
\hline
 & $\mathrm{LO}+\mathrm{NNLL}$ &
$\mathrm{N}^3\mathrm{LO}_\mathrm{approx}+\mathrm{NNLL}$ &
$\mathrm{N}^3\mathrm{LO}_\mathrm{approx}+\mathrm{NLO}+\mathrm{NNLL}$ \\
\hline
$\langle n_h(Q_0^2)\rangle_g$ & $24.31\pm0.85$ & $24.02\pm0.36$ &
$24.17\pm 0.36$ \\
$\langle n_h(Q_0^2)\rangle_q$ & $15.49\pm0.90$ & $15.83\pm0.37$ &
$15.89\pm 0.33$ \\
$\chi_\mathrm{dof}^2$ & 18.09 & 3.71 & 2.92 \\
\hline
\end{tabular}
\caption{\footnotesize%
Fit results for $\langle n_h(Q_0^2)\rangle_g$ and $\langle n_h(Q_0^2)\rangle_q$
at $Q_0=50$~GeV with 90\% CL errors and minimum values of
$\chi_\mathrm{dof}^2$ achieved in the $\mathrm{LO}+\mathrm{NNLL}$,
$\mathrm{N}^3\mathrm{LO}_\mathrm{approx}+\mathrm{NNLL}$, and 
$\mathrm{N}^3\mathrm{LO}_\mathrm{approx}+\mathrm{NLO}+\mathrm{NNLL}$
approximations.}
\label{tab:fit}
\end{table}

The average gluon and quark jet multiplicities extracted from experimental
data strongly depend on the choice of jet algorithm.
We adopt the selection of experimental data from Ref.~\cite{Abdallah:2005cy}
performed in such a way that they correspond to compatible jet algorithms.
Specifically, these include the measurements of average gluon jet
multiplicities in
Refs.~\cite{Abdallah:2005cy,Nakabayashi:1997hr,Abbiendi:1999pi,%
Abbiendi:2004pr,Siebel:2003zz}
and those of average quark jet multiplicities in
Refs.~\cite{Nakabayashi:1997hr,%
Kluth:2003uq,Althoff:1983ew,Braunschweig:1989bp,Aihara:1986mv,%
Rowson:1985xh,Derrick:1986jx,Zheng:1990iq,Abrams:1989rz,%
Decamp:1989tf,Decamp:1991uz,Buskulic:1995xz,Barate:1996fi,Abreu:1990cc,%
Abreu:1991yc,Adeva:1991it,Adeva:1992gv,Akrawy:1990yx,Acton:1992ry,%
Acton:1991aa,Abreu:1998vq,Ackerstaff:1998hz,Acciarri:1995ia,Abreu:1996va,%
Alexander:1996kh,Buskulic:1996tt,Ackerstaff:1997kk,Abreu:1997dm,%
Abbiendi:1999sx,Abreu:2000gw}, which include 27 and 51 experimental data
points, respectively.
The results for $\langle n_h(Q_0^2)\rangle_g$ and
$\langle n_h(Q_0^2)\rangle_q$ at $Q_0=50$~GeV together with the
$\chi_\mathrm{dof}^2$ values obtained in our $\mathrm{LO}+\mathrm{NNLL}$,
$\mathrm{N}^3\mathrm{LO}_\mathrm{approx}+\mathrm{NNLL}$, and 
$\mathrm{N}^3\mathrm{LO}_\mathrm{approx}+\mathrm{NLO}+\mathrm{NNLL}$ fits are
listed in Table~\ref{tab:fit}.
The errors correspond to 90\% CL as explained above.
All these fit results are in agreement with the experimental data.
Looking at the $\chi_\mathrm{dof}^2$ values, we observe that the qualities of
the fits improve as we go to higher orders, as they should.
The improvement is most dramatic in the step from
$\mathrm{LO}+\mathrm{NNLL}$ to
$\mathrm{N}^3\mathrm{LO}_\mathrm{approx}+\mathrm{NNLL}$, where the errors on
$\langle n_h(Q_0^2)\rangle_g$ and $\langle n_h(Q_0^2)\rangle_q$ are more than
halved.
The improvement in the step from
$\mathrm{N}^3\mathrm{LO}_\mathrm{approx}+\mathrm{NNLL}$ to 
$\mathrm{N}^3\mathrm{LO}_\mathrm{approx}+\mathrm{NLO}+\mathrm{NNLL}$, albeit
less pronounced, indicates that the inclusion of the first correction to
$r_-(Q^2)$ as given in Eq.~(\ref{frminus}) is favored by the experimental data.
We have verified that the values of $\chi_\mathrm{dof}^2$ are insensitive to
the choice of $Q_0$, as they should. 
Furthermore, the central values converge in the sense that the shifts in the
step from $\mathrm{N}^3\mathrm{LO}_\mathrm{approx}+\mathrm{NNLL}$ to 
$\mathrm{N}^3\mathrm{LO}_\mathrm{approx}+\mathrm{NLO}+\mathrm{NNLL}$ are
considerably smaller than those in the step from $\mathrm{LO}+\mathrm{NNLL}$ to
$\mathrm{N}^3\mathrm{LO}_\mathrm{approx}+\mathrm{NNLL}$ and that, at the same
time, the central values after each step are contained within error bars before
that step.
In the fits presented so far, the strong-coupling constant was taken to be the
central value of the world avarage, $\alpha_s^{(5)}(m_Z^2)=0.1184$
\cite{Beringer:1900zz}.
In Section~\ref{coupling}, we shall include $\alpha_s^{(5)}(m_Z^2)$ among the
fit parameters.

\begin{figure}
\centering
\includegraphics[width=0.85\textwidth]{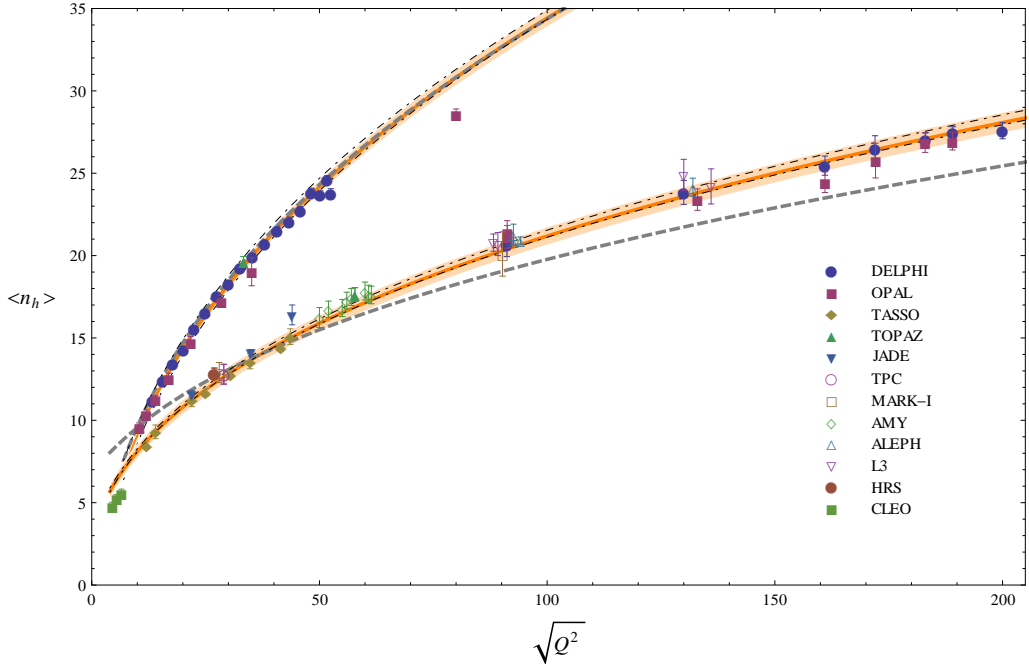}
\caption{\footnotesize{%
The average gluon (upper curves) and quark (lower curves) jet multiplicities
evaluated from Eq.~(\ref{quarkgen}), respectively, in the
$\mathrm{LO}+\mathrm{NNLL}$ (dashed/gray lines) and
$\mathrm{N}^3\mathrm{LO}_\mathrm{approx}+\mathrm{NLO}+\mathrm{NNLL}$ 
(solid/orange lines) approximations using the corresponding fit results for
$\langle n_h(Q_0^2)\rangle_g$ and $\langle n_h(Q_0^2)\rangle_q$ from
Table~\ref{tab:fit} are compared with the experimental data included in the
fits.
The experimental and theoretical uncertainties in the
$\mathrm{N}^3\mathrm{LO}_\mathrm{approx}+\mathrm{NLO}+\mathrm{NNLL}$ results
are indicated by the shaded/orange bands and the bands enclosed between the
dot-dashed curves, respectively.}}
\label{Fig:plotmult}
\end{figure}

\begin{figure}
\centering
\includegraphics[width=0.85\textwidth]{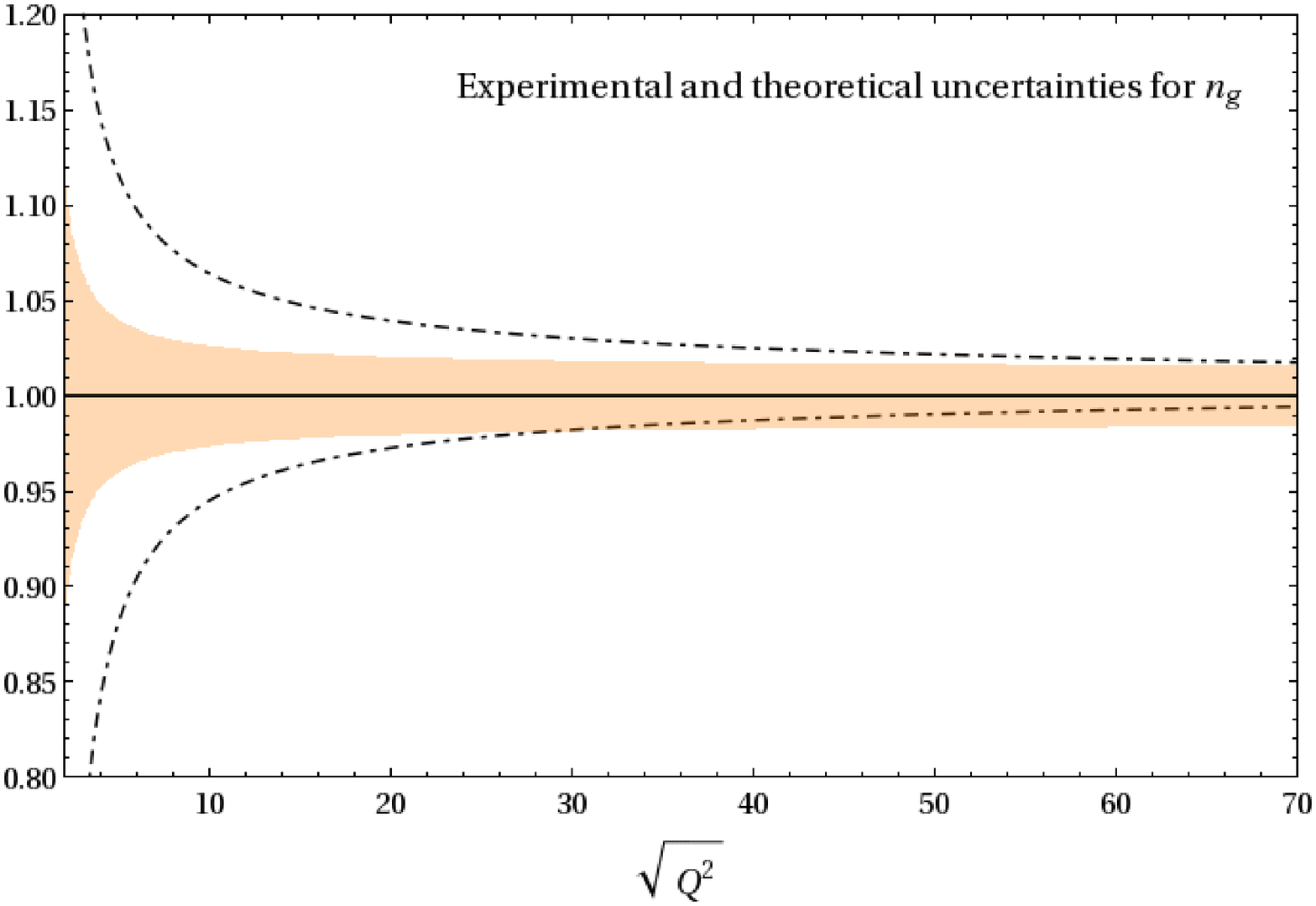}
\caption{\footnotesize{%
Experimental (shaded/orange band) and theoretical (band enclosed between
dot-dashed curves) uncertainties in the
$\mathrm{N}^3\mathrm{LO}_\mathrm{approx}+\mathrm{NLO}+\mathrm{NNLL}$ result
for the average gluon jet multiplicity normalized with respect to default
evaluation with $\xi=1$.}}
\label{Fig:gluon_unc}
\end{figure}

\begin{figure}
\centering
\includegraphics[width=0.85\textwidth]{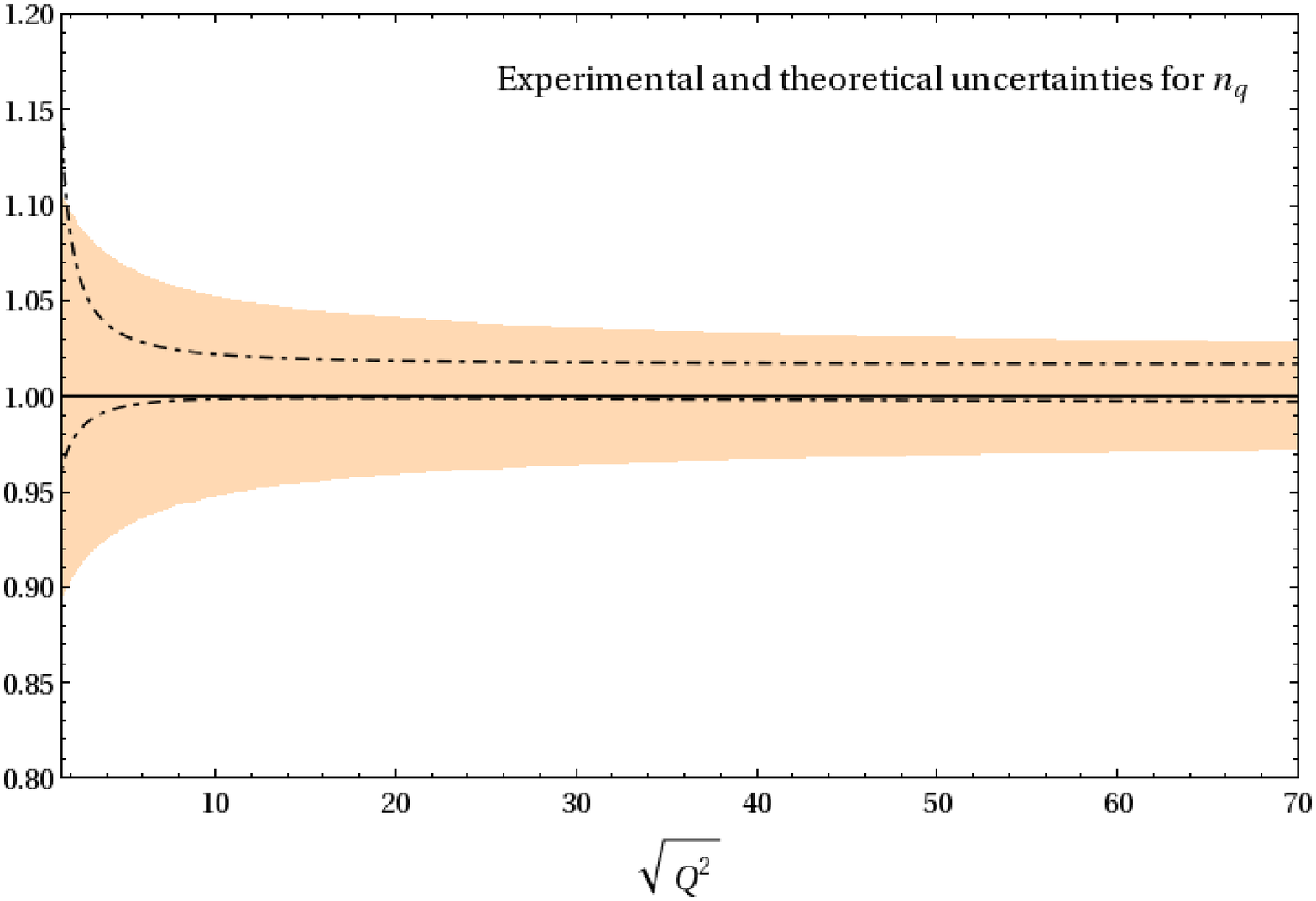}
\caption{\footnotesize{%
Experimental (shaded/orange band) and theoretical (band enclosed between
dot-dashed curves) uncertainties in the
$\mathrm{N}^3\mathrm{LO}_\mathrm{approx}+\mathrm{NLO}+\mathrm{NNLL}$ result
for the average quark jet multiplicity normalized with respect to default
evaluation with $\xi=1$.}}
\label{Fig:quark_unc}
\end{figure}

In Fig.~\ref{Fig:plotmult}, we show as functions of $Q$ the average gluon and
quark jet multiplicities evaluated from Eq.~(\ref{quarkgen}) at
$\mathrm{LO}+\mathrm{NNLL}$ and 
$\mathrm{N}^3\mathrm{LO}_\mathrm{approx}+\mathrm{NLO}+\mathrm{NNLL}$ using the
corresponding fit results for $\langle n_h(Q_0^2)\rangle_g$ and
$\langle n_h(Q_0^2)\rangle_q$ at $Q_0=50$~GeV from Table~\ref{tab:fit}.
For clarity, we refrain from including in Fig.~\ref{Fig:plotmult} the
$\mathrm{N}^3\mathrm{LO}_\mathrm{approx}+\mathrm{NNLL}$ results, which are very
similar to the 
$\mathrm{N}^3\mathrm{LO}_\mathrm{approx}+\mathrm{NLO}+\mathrm{NNLL}$ ones
already presented in Ref.~\cite{Bolzoni:2012ii}.
In the $\mathrm{N}^3\mathrm{LO}_\mathrm{approx}+\mathrm{NLO}+\mathrm{NNLL}$
case, Fig.~\ref{Fig:plotmult} also displays two error bands, namely the
experimental one induced by the 90\% CL errors on the respective fit parameters
in Table~\ref{tab:fit} and the theoretical one, which is evaluated from
Eqs.~(\ref{shiftA}) and (\ref{dreminscale}) by varying the scale parameter
$\xi$ in the range $1/4\le\xi\le4$.
For a more detailed discussion of the uncertainties on the average gluon and
quark jet multiplicities in the
$\mathrm{N}^3\mathrm{LO}_\mathrm{approx}+\mathrm{NLO}+\mathrm{NNLL}$
approximation, we display them as functions of $Q$ in Fig.~\ref{Fig:gluon_unc}
and \ref{Fig:quark_unc}, respectively, normalized with respect to the default
results, evaluated with $\xi=1$.
We observe that the uncertainties decrease with increasing value of $Q$, which
is a consequence of the asymptotic freedom of QCD.
They typically fall below $\pm5\%$ at $Q\approx10$~GeV, but become significant
at low $Q$ values indicating the onset of the breakdown of the perturbative
expansion in $\sqrt{\alpha_s}$.

\begin{figure}
\centering
\includegraphics[width=0.85\textwidth]{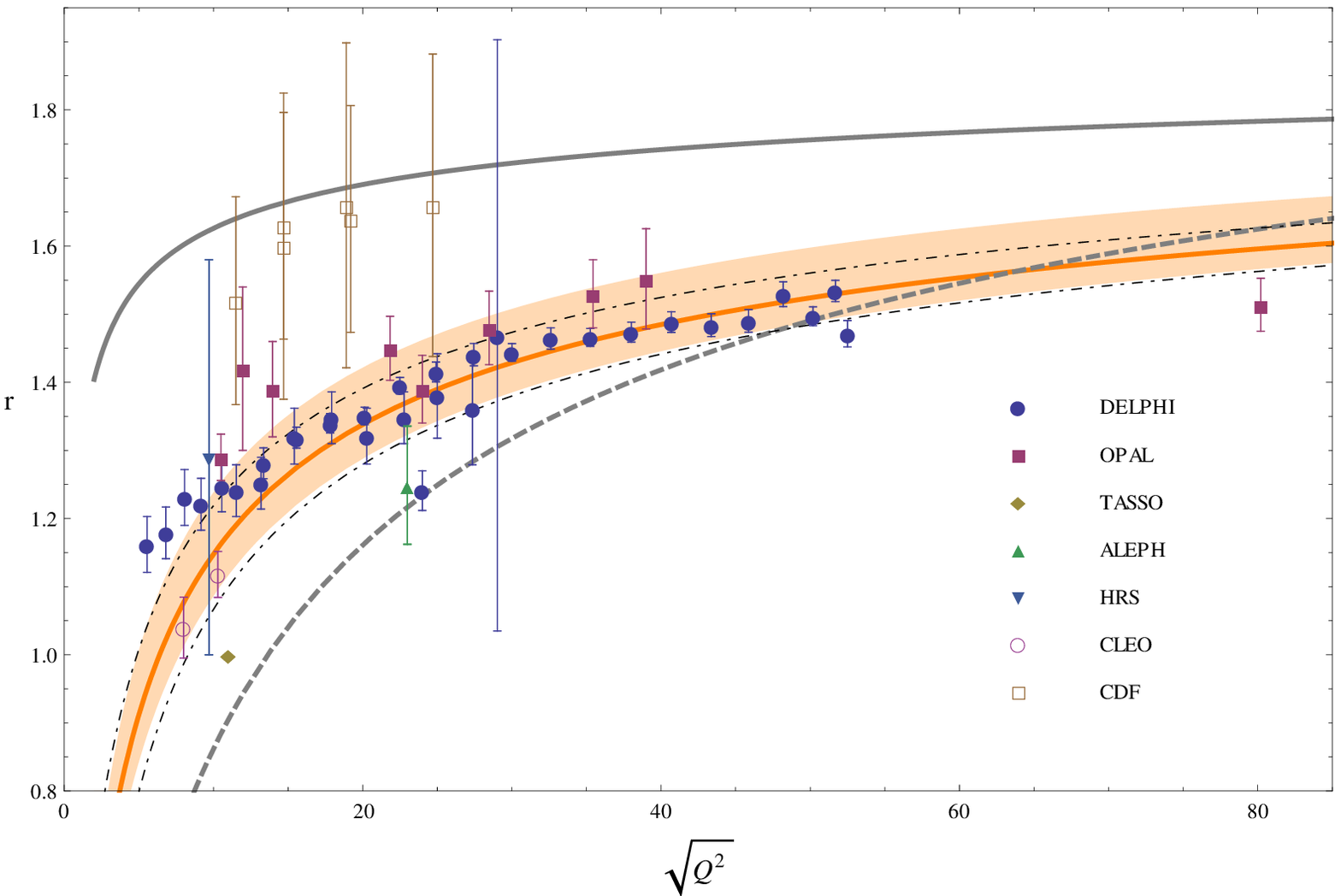}
\caption{\footnotesize{%
The average gluon-to-quark jet multiplicity ratio evaluated from
Eq.~(\ref{ratiogen}) in the $\mathrm{LO}+\mathrm{NNLL}$ (dashed/gray lines) and 
$\mathrm{N}^3\mathrm{LO}_\mathrm{approx}+\mathrm{NLO}+\mathrm{NNLL}$ 
(solid/orange lines) approximations using the corresponding fit results for
$\langle n_h(Q_0^2)\rangle_g$ and $\langle n_h(Q_0^2)\rangle_q$ from
Table~\ref{tab:fit} are compared with experimental data.
The experimental and theoretical uncertainties in the
$\mathrm{N}^3\mathrm{LO}_\mathrm{approx}+\mathrm{NLO}+\mathrm{NNLL}$ result are
indicated by the shaded/orange bands and the bands enclosed between the
dot-dashed curves, respectively.
The prediction given by Eq.~(\ref{dreminscaleplus}) \cite{Capella:1999ms} is
indicated by the continuous/gray line.}}
\label{Fig:ratio}
\end{figure}

\begin{figure}
\centering
\includegraphics[width=0.85\textwidth]{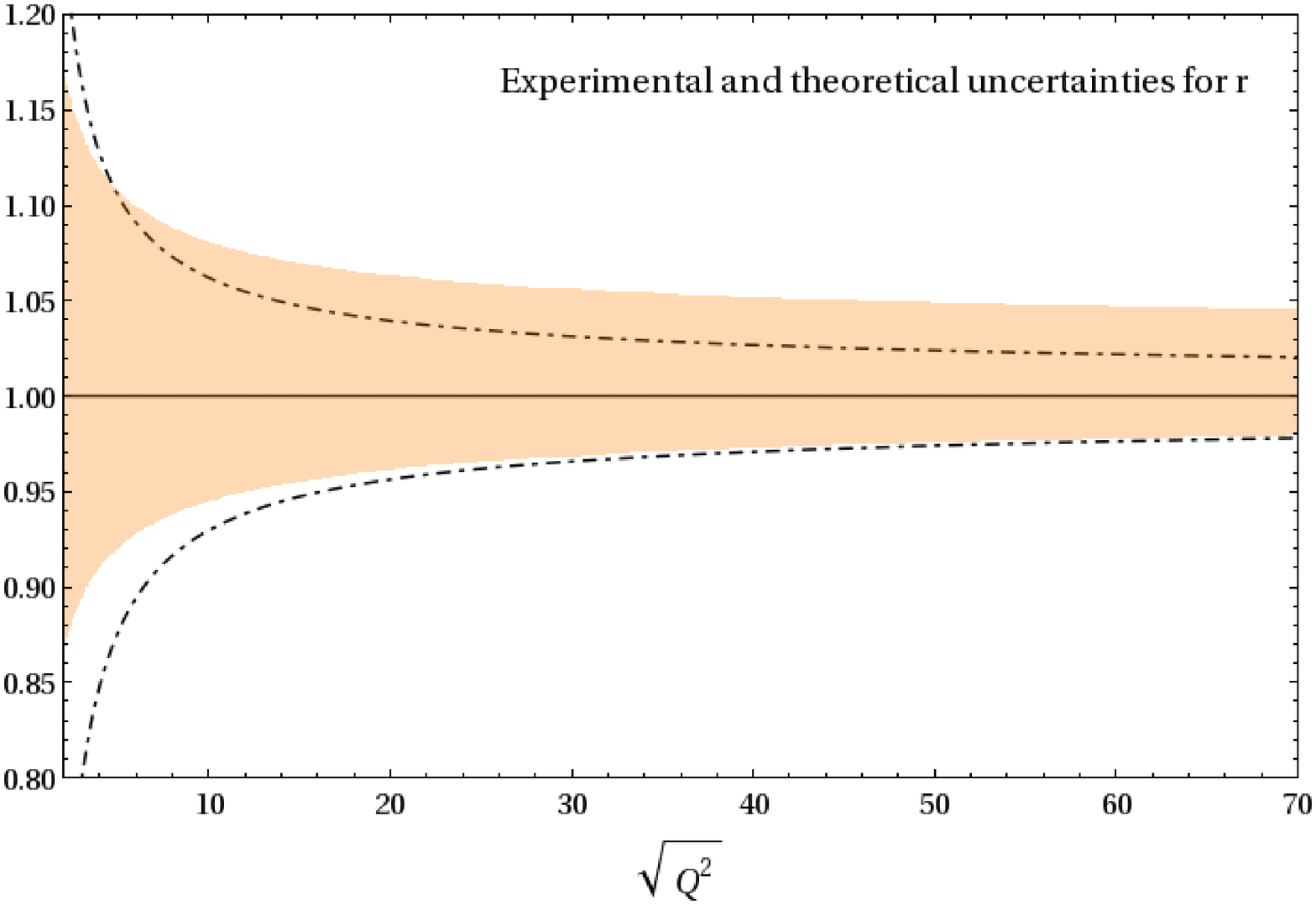}
\caption{\footnotesize{%
Experimental (dark/orange band) and theoretical (light/gray band) uncertainties
in the $\mathrm{N}^3\mathrm{LO}_\mathrm{approx}+\mathrm{NLO}+\mathrm{NNLL}$
result for the average gluon-to-quark jet multiplicity ratio normalized with
respect to default evaluation with $\xi=1$.}}
\label{Fig:ratio_unc}
\end{figure}

While our fits rely on individual measurements of the average gluon and quark
jet multiplicities, the experimental literature also reports determinations of
their ratio; see
Refs.~\cite{Abreu:1999rs,Abdallah:2005cy,Abbiendi:1999pi,Siebel:2003zz,%
Alam:1997ht,Albrecht:1991vp,Alam:1992ir,Acosta:2004js,Derrick:1985du,%
Braunschweig:1989um,Alexander:1991ce,Acton:1993jm,OPAL:1995ab,Biebel:1996mc,%
Buskulic:1995sw,Abreu:1995hp,Alexander:1996qr,Ackerstaff:1997xg,%
Abbiendi:2003gh}, which essentially cover all the available measurements.
In order to find out how well our fits describe the latter and thus to test
the global consistency of the individual measurements, we compare in
Fig.~\ref{Fig:ratio} the experimental data on the average gluon-to-quark jet
multiplicity ratio with our evaluations of Eq.~(\ref{ratiogen}) in the
$\mathrm{LO}+\mathrm{NNLL}$ and 
$\mathrm{N}^3\mathrm{LO}_\mathrm{approx}+\mathrm{NLO}+\mathrm{NNLL}$
approximations using the corresponding fit results from Table~\ref{tab:fit}.
As in Fig.~\ref{Fig:plotmult}, we present in Fig.~\ref{Fig:ratio} also the
experimental and theoretical uncertainties in the
$\mathrm{N}^3\mathrm{LO}_\mathrm{approx}+\mathrm{NLO}+\mathrm{NNLL}$ result.
As in Figs.~\ref{Fig:gluon_unc} and \ref{Fig:quark_unc}, they are
represented relative to the default result, with $\xi=1$, in
Fig.~\ref{Fig:ratio_unc}.
For comparison, we include in Fig.~\ref{Fig:ratio} also the prediction of
Ref.~\cite{Capella:1999ms} given by Eq.~(\ref{dreminscaleplus}).

Looking at Fig.~\ref{Fig:ratio}, we observe that the experimental data are
very well described by the 
$\mathrm{N}^3\mathrm{LO}_\mathrm{approx}+\mathrm{NLO}+\mathrm{NNLL}$ result for
$Q$ values above 10~GeV, while they somewhat overshoot it below.
This discrepancy is likely to be due to the fact that, following
Ref.~\cite{Abdallah:2005cy}, we excluded the older data from 
Ref.~\cite{Abreu:1999rs} from our fits because they are inconsistent with the
experimental data sample compiled in Ref.~\cite{Abdallah:2005cy}.
Furthermore, Fig.~\ref{Fig:ratio_unc} tells us that the theoretical
uncertainties are large in the small-$Q^2$ region, which indicates that the
convergence properties of the perturbative series in $\sqrt{\alpha_s}$ are
unfavorable there.
Finally, the experimental determination of the scale $Q$, which in the
theoretical expressions denotes the virtuality of the parent parton of the
considered jet, may be ambiguous in multi-jet events and may be performed
somewhat differently in different experiments, which may explain tensions
between different data sets.
This additional type of uncertainty should be more important at small values of
$Q^2$, where the slope of the $Q^2$ evolution is steeper.

The Monte Carlo analysis of Ref.~\cite{Eden:1998ig} suggests that the average
gluon and quark jet multiplicities should coincide at about $Q=4$~GeV.
As is evident from Fig.~\ref{Fig:ratio}, this agrees with our
$\mathrm{N}^3\mathrm{LO}_\mathrm{approx}+\mathrm{NLO}+\mathrm{NNLL}$ result
reasonably well given the considerable uncertainties in the small-$Q^2$ range
discussed above.

\begin{figure}
\centering
\includegraphics[width=0.85\textwidth]{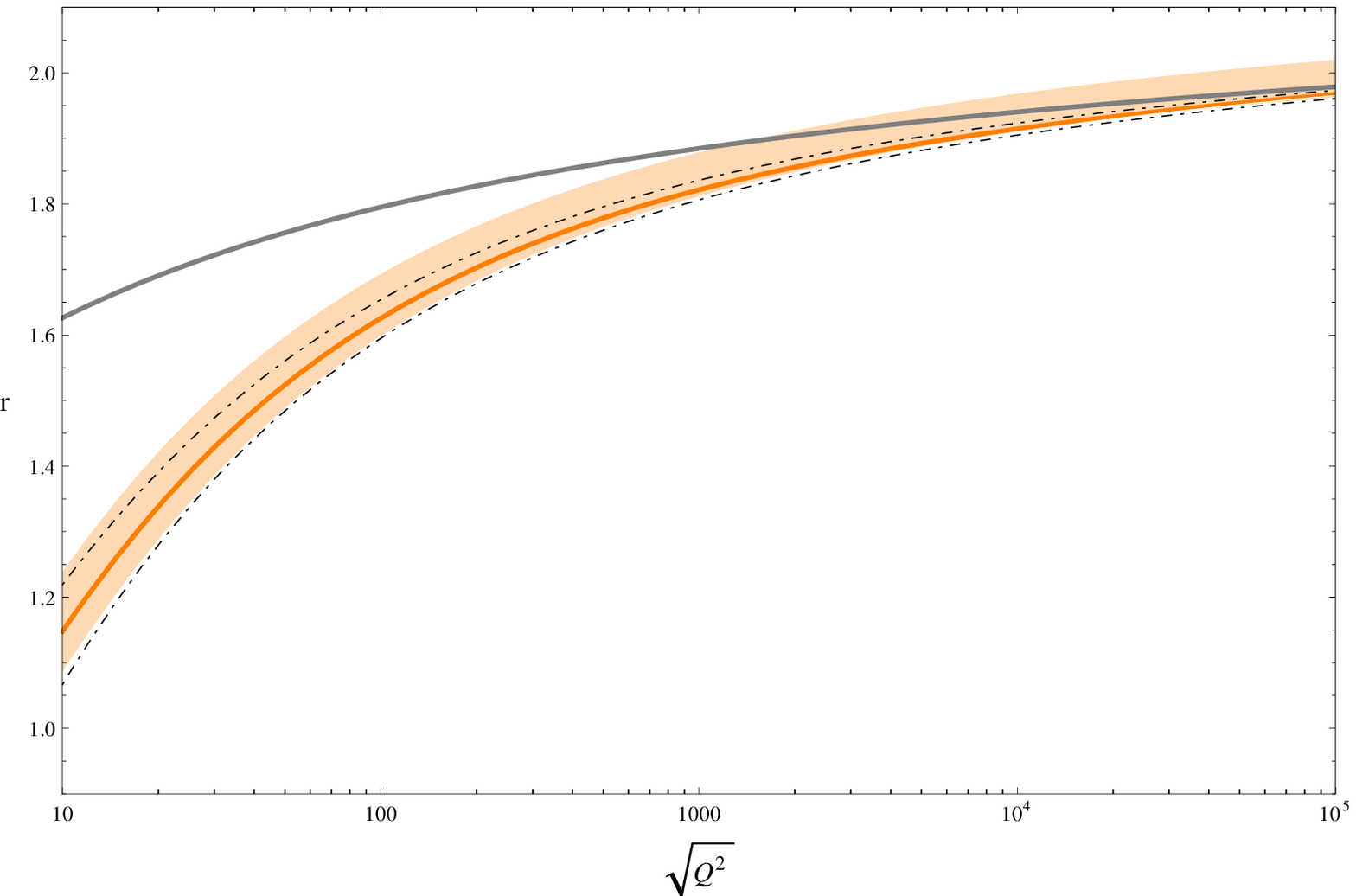}
\caption{\footnotesize{%
High-$Q$ extension of Fig.~\ref{Fig:ratio}.}}
\label{Fig:high_en_ratio}
\end{figure}

As is obvious from Fig.~\ref{Fig:ratio}, the approximation of $r(Q^2)$ by
$r_+(Q^2)$ given in Eq.~(\ref{dreminscaleplus}) \cite{Capella:1999ms} leads to
a poor approximation of the experimental data, which reach up to $Q$ values of
about 50~GeV.
It is, therefore, interesting to study the high-$Q^2$ asymptotic behavior of
the average gluon-to-quark jet ratio.
This is done in Fig.~\ref{Fig:high_en_ratio}, where the
$\mathrm{N}^3\mathrm{LO}_\mathrm{approx}+\mathrm{NLO}+\mathrm{NNLL}$ result
including its experimental and theoretical uncertainties is compared with the
approximation by Eq.~(\ref{dreminscaleplus}) way up to $Q=100$~TeV.
We observe from Fig.~\ref{Fig:high_en_ratio} that the approximation
approaches the
$\mathrm{N}^3\mathrm{LO}_\mathrm{approx}+\mathrm{NLO}+\mathrm{NNLL}$ result
rather slowly.
Both predictions agree within theoretical errors at $Q=100$~TeV, which is one
order of magnitude beyond LHC energies, where they are still about 10\% below
the asymptotic value $C_A/C_F=2.25$.
Figure~\ref{Fig:high_en_ratio} also nicely illustrates how, as a consequence of
the asymptotic freedom of QCD, the theoretical uncertainty decreases with
increasing value of $Q^2$ and thus becomes considerably smaller than the
experimental error.

\section{Determination of strong-coupling constant}
\label{coupling}

\begin{table}
\centering
\begin{tabular}{|c|c|c|}
\hline
 & $\mathrm{N}^3\mathrm{LO}_\mathrm{approx}+\mathrm{NNLL}$ &
$\mathrm{N}^3\mathrm{LO}_\mathrm{approx}+\mathrm{NLO}+\mathrm{NNLL}$ \\
\hline
$\langle n_h(Q_0^2)\rangle_g$ & $24.18\pm0.32$ & $24.22\pm 0.33$ \\
$\langle n_h(Q_0^2)\rangle_q$ & $15.86\pm0.37$ & $15.88\pm 0.35$ \\
$\alpha_s^{(5)}(m_Z^2)$ & $0.1242\pm0.0046$ & $0.1199\pm0.0044$ \\
$\chi_\mathrm{dof}^2$ & 2.84 & 2.85 \\
\hline
\end{tabular}
\caption{\footnotesize%
Fit results for $\langle n_h(Q_0^2)\rangle_g$ and $\langle n_h(Q_0^2)\rangle_q$
at $Q_0=50$~GeV and for $\alpha_s^{(5)}(m_Z^2)$ with 90\% CL errors and minimum
values of $\chi_\mathrm{dof}^2$ achieved in the
$\mathrm{N}^3\mathrm{LO}_\mathrm{approx}+\mathrm{NNLL}$ and 
$\mathrm{N}^3\mathrm{LO}_\mathrm{approx}+\mathrm{NLO}+\mathrm{NNLL}$
approximations.}
\label{tab:fit2}
\end{table}

In Section~\ref{analysis}, we took $\alpha_s^{(5)}(m_Z^2)$ to be a fixed input
parameter for our fits.
Motivated by the excellent goodness of our
$\mathrm{N}^3\mathrm{LO}_\mathrm{approx}+\mathrm{NNLL}$ and
$\mathrm{N}^3\mathrm{LO}_\mathrm{approx}+\mathrm{NLO}+\mathrm{NNLL}$ fits, we
now include it among the fit parameters, the more so as the fits should be
sufficiently sensitive to it in view of the wide $Q^2$ range populated by the
experimental data fitted to.
We fit to the same experimental data as before and again put $Q_0=50$~GeV. 
The fit results are summarized in Table~\ref{tab:fit2}.
We observe from Table~\ref{tab:fit2} that the results of the
$\mathrm{N}^3\mathrm{LO}_\mathrm{approx}+\mathrm{NNLL}$ \cite{Bolzoni:2012cv}
and $\mathrm{N}^3\mathrm{LO}_\mathrm{approx}+\mathrm{NLO}+\mathrm{NNLL}$ fits
for $\langle n_h(Q_0^2)\rangle_g$ and $\langle n_h(Q_0^2)\rangle_q$ are
mutually consistent.
They are also consistent with the respective fit results in
Table~\ref{tab:fit}.
As expected, the values of $\chi_\mathrm{dof}^2$ are reduced by relasing
$\alpha_s^{(5)}(m_Z^2)$ in the fits, from 3.71 to 2.84 in the
$\mathrm{N}^3\mathrm{LO}_\mathrm{approx}+\mathrm{NNLL}$ approximation and
from 2.95 to 2.85 in the
$\mathrm{N}^3\mathrm{LO}_\mathrm{approx}+\mathrm{NLO}+\mathrm{NNLL}$ one.
The three-parameter fits strongly confine $\alpha_s^{(5)}(m_Z^2)$, within an
error of 3.7\% at 90\% CL in both approximations.
The inclusion of the $r_-(Q^2)$ term has the beneficial effect of shifting
$\alpha_s^{(5)}(m_Z^2)$ closer to the world average, $0.1184\pm0.0007$
\cite{Beringer:1900zz}.
In fact, our
$\mathrm{N}^3\mathrm{LO}_\mathrm{approx}+\mathrm{NLO}+\mathrm{NNLL}$ value,
$0.1199\pm0.0044$ at 90\% CL, which corresponds to $0.1199\pm0.0026$ at 68\%
CL, is in excellent agreement with the former.

\begin{figure}
\centering
\includegraphics[width=0.85\textwidth]{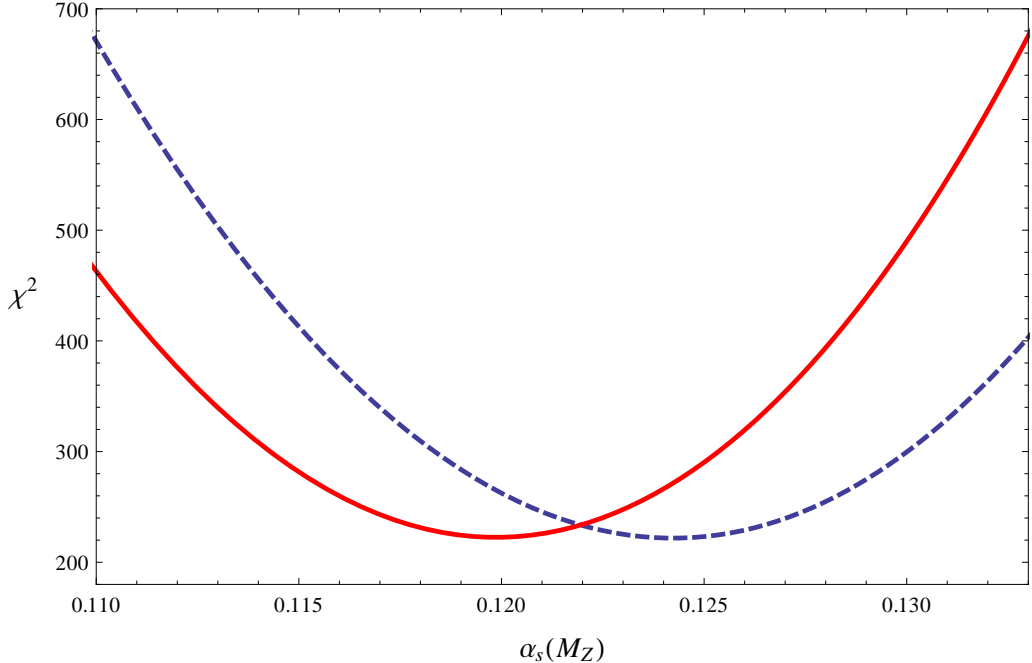}
\caption{\footnotesize{%
Values of $\chi^2$ evaluated as functions of $\alpha_s^{(5)}(m_Z^2)$ in the
$\mathrm{N}^3\mathrm{LO}_\mathrm{approx}+\mathrm{NNLL}$ and
$\mathrm{N}^3\mathrm{LO}_\mathrm{approx}+\mathrm{NLO}+\mathrm{NNLL}$
approximations with the respective central values of
$\langle n_h(Q_0^2)\rangle_g$ and $\langle n_h(Q_0^2)\rangle_q$ from
Table~\ref{tab:fit2}.}}
\label{Fig:as_parabola}
\end{figure}

In order to illustrate the sensitivity of our
$\mathrm{N}^3\mathrm{LO}_\mathrm{approx}+\mathrm{NNLL}$ and
$\mathrm{N}^3\mathrm{LO}_\mathrm{approx}+\mathrm{NLO}+\mathrm{NNLL}$ fits to
$\alpha_s^{(5)}(m_Z^2)$, we show in Fig.~\ref{Fig:as_parabola} the values of
$\chi^2$ obtained by varying $\alpha_s^{(5)}(m_Z^2)$ while keeping
$\langle n_h(Q_0^2)\rangle_g$ and $\langle n_h(Q_0^2)\rangle_q$ at their
respective central values listed in Table~\ref{tab:fit2}.

\section{Conclusions}
\label{conclusions}

Prior to our analysis in Ref.~\cite{Bolzoni:2012ii}, experimental data on the
average gluon and quark jet multiplicities could not be simultaneously
described in a satisfactory way mainly because the theoretical formalism failed
to account for the difference in hadronic contents between gluon and quark
jets, although the convergence of perturbation theory seemed to be well under
control \cite{Capella:1999ms}.
This problem may be solved by including the minus components governed by
$\hat{T}_-^\mathrm{res}(0,Q^2,Q_0^2)$ in Eqs.~(\ref{quarkgen}) and
(\ref{ratiogen}).
This was done for the first time in Ref.~\cite{Bolzoni:2012ii}, albeit in
connection with the LO result $r_-(Q^2)=0$.
The quark-singlet minus component comes with an arbitrary normalization and has
a slow $Q^2$ dependence.
Consequently, its numerical contribution may be approximately mimicked by a
constant introduced to the average quark jet multiplicity as in
Ref.~\cite{Abreu:1999rs}.

In the present paper, we improved the analysis of Ref.~\cite{Bolzoni:2012ii} in
various ways.
The most natural possible improvement consists in including higher-order
correction to $r_-(Q^2)$.
Here, we managed to obtain the NLO correction, of
$\mathcal{O}(\sqrt{\alpha_s})$, using the effective approach introduced in
Ref.~\cite{Bolzoni:2012ed}, which was shown to also exactly reproduce the
$\mathcal{O}(\sqrt{\alpha_s})$ correction to $r_+(Q^2)$.
Our general result corresponding to Eq.~(\ref{quarkgen}) 
depends on two parameters, $D_g(0,Q_0^2)$ and $D_s(0,Q_0^2)$, which,
according to Eq.~(\ref{incond}), represent the average gluon and quark jet
multiplicities at an arbitrary reference scale $Q_0$ and act as initial
conditions for the $Q^2$ evolution.
Looking at the perturbative behaviour of the expansion in $\sqrt{\alpha_s}$
and the distribution of the available experimental data, we argued that
$Q_0=50$~GeV is a good choice.
We fitted these two parameters to all available experimental data on the
average gluon and quark jet multiplicities treating $\alpha_s^{(5)}(m_Z^2)$ as
an input parameter fixed to the world avarage \cite{Beringer:1900zz}.
We worked in three different approximations, labeled
$\mathrm{LO}+\mathrm{NNLL}$,
$\mathrm{N}^3\mathrm{LO}_\mathrm{approx}+\mathrm{NNLL}$, and
$\mathrm{N}^3\mathrm{LO}_\mathrm{approx}+\mathrm{NLO}+\mathrm{NNLL}$,
in which the logarithms $\ln x$ are resummed through the NNLL level,
$r_+(Q^2)$ is evaluated at LO or approximately at N$^3$LO, and $r_-(Q^2)$ is
evaluated at LO or NLO.
Including the NLO correction to $r_-(Q^2)$, given in Eq.~(\ref{frminus}),
significantly improved the quality of the fit, as is evident by comparing the
values of $\chi_\mathrm{dof}^2$ for the
$\mathrm{N}^3\mathrm{LO}_\mathrm{approx}+\mathrm{NNLL}$ and
$\mathrm{N}^3\mathrm{LO}_\mathrm{approx}+\mathrm{NLO}+\mathrm{NNLL}$ fits in
Table~\ref{tab:fit}.

Motivated by the goodness of our
$\mathrm{N}^3\mathrm{LO}_\mathrm{approx}+\mathrm{NNLL}$ and
$\mathrm{N}^3\mathrm{LO}_\mathrm{approx}+\mathrm{NLO}+\mathrm{NNLL}$ fits with
fixed value of $\alpha_s^{(5)}(m_Z^2)$ in Ref.~\cite{Bolzoni:2012ii} and here,
we then included $\alpha_s^{(5)}(m_Z^2)$ among the fit parameters, which
yielded a further reduction of $\chi_\mathrm{dof}^2$.
The fit results are listed in Table~\ref{tab:fit2}.
Also here, the inclusion of the NLO correction to $r_-(Q^2)$ is beneficial;
it shifts $\alpha_s^{(5)}(m_Z^2)$ closer to the world average to become
$0.1199\pm0.0026$.

A few comments are in order regarding the renormalization scheme and the
counting of higher-order corrections in our analysis in order to allow for an
appropriate classification of our determination of $\alpha_s^{(5)}(m_Z^2)$ in
the context of a global analysis yielding a world average.
We worked in the $\overline{\mathrm{MS}}$ renormalization scheme, which has
become the standard choice in the literature.
We reach beyond ordinary fixed-order analyses by resumming the logarithms
$\ln x$ through the NNLL level.
Furthermore, our expressions are completely RG-improved in the sense that all
$Q^2$ dependence is accommodated in $\alpha_s(Q^2)$.
Unlike usual higher-order calculations in the QCD-improved parton model, the
perturbation series of the coefficients $r_\pm(Q^2)$ are organized in powers of
$\sqrt{\alpha_s}$ rather than $\alpha_s$.
In the case of $r_+(Q^2)$, which starts at $\mathrm{O}(1)$, our exact knowledge
reaches through $\mathrm{O}(\alpha_s)$, i.e.\ NNLO, while our
$\mathrm{O}(\alpha_s^{3/2})$ term represents an educated guess in the sense
that it was obtained using a procedure that, strictly speaking, was only tested
through NNLO.
In the case of $r_-(Q^2)$, the $\mathrm{O}(1)$ term vanishes, and the
$\mathrm{O}(\sqrt{\alpha_s})$ term is listed in Eq.~(\ref{frminus}), i.e.\
we have control through NLO.
However, the coefficients of $r_-(Q^2)$ in Eq.~(\ref{quarkgen}) are numerically
suppressed relative to those of $r_+(Q^2)$, by approximately a factor of
$\mathrm{O}(\sqrt{\alpha_s})$.
In fact, the shift in $\langle n_h(Q^2)\rangle_g$ ($\langle n_h(Q^2)\rangle_s$)
induced by the $\mathrm{O}(\sqrt{\alpha_s})$ term of $r_-(Q^2)$ is comparable
to (about a factor of three smaller than) the one induced by the
$\mathrm{O}(\alpha_s)$ term of $r_+(Q^2)$.
We thus conclude that our determination of $\alpha_s^{(5)}(m_Z^2)$ is effectively
of NNLO.

The next steps towards $\mathcal{O}(\alpha_s^2)$ accuracy include an improved
computation of the coefficient $r_-(Q^2)$ and an extended resummation of the
plus and minus components of the splitting functions.
At the LHC, jet multiplicity observables can be measured at unprecedented
values of $Q^2$, which will allow for stringent tests of QCD and provide a
strong lever arm for high-precision determinations of $\alpha_s^{(5)}(m_Z^2)$
using the formalism elaborated in Ref.~\cite{Bolzoni:2012ii} and here.

\subsection*{Acknowledgments}

The work P.B. and of A.V.K. was supported in part by the Heisenberg-Landau
program.
The work of A.V.K. was supported in part by the Russian Foundation for Basic
Research RFBR through Grant No.\ 13-02-01005.
This work was supported by the German Federal Ministry for Education
and Research BMBF through Grant No.\ 05H12GUE and by the German Research
Foundation DFG through the Collaborative Research Centre No.~676
{\it Particles, Strings and the Early Universe---The Structure of Matter and
Space Time}.

\begin{appendix}

\section*{Appendix}
\label{appendix}

Here we prove the relations given in Eqs.~(\ref{tolja3}) and (\ref{tolja3.1}) 
between the singular parts of the diagonal and nondiagonal splitting functions
in Mellin space $P_{ab}(\omega,a_s)$ with $a,b=g,q$ and show that they are
{\it approximately} true also for the regular parts.

Following Ref.~\cite{Kom:2012hd}, we introduce the notation\footnote[2]{%
In order for all variables to be positive, we introduce here $\eta$ instead of
$\xi$ used in Ref.~\cite{Kom:2012hd}.}
\begin{equation}
\eta = \frac{8 C_A a_s}{\omega^2},\qquad
s=\sqrt{1+4\eta},\qquad
L=\ln\frac{1+s}{2}=\ln\frac{2\eta}{s-1}.
\end{equation}
In the following, the resummed functions $P_{ab}(\omega,a_s)$ are built up by
their parts $P_{ab}^{(i)}(\omega)$ corresponding to the considered levels of
resummation, with $i=0,1,2$ representing the LL, NLL, and NNLL levels,
respectively.
The results read:
\begin{eqnarray}
P_{qq}(\omega,a_s) &=& P_{qq}^{(1)}(\omega,a_s) + P_{qq}^{(2)}(\omega,a_s),
\nonumber\\
P_{qq}^{(1)}(\omega,a_s)&=&\frac{4}{3} C_F f_A a_s
\left[ 1-\frac{s-1}{2\eta} (L+1)\right],
\nonumber\\
P_{qq}^{(2)}(\omega,a_s)&=& C_F f_A a_s K^{qq}
\nonumber\\
P_{gg}(\omega,a_s)&=&-P_{qq}(\omega,a_s)
+\tilde{P}_{gg}^{(0)}(\omega,a_s)+ \tilde{P}_{gg}^{(1)}(\omega,a_s) 
+ \tilde{P}_{qq}^{(2)}(\omega,a_s),
\nonumber\\
\tilde{P}_{gg}^{(0)}(\omega,a_s) &=& \frac{\omega}{4}(s-1),
\nonumber\\
\tilde{P}_{gg}^{(1)}(\omega,a_s) &=& \frac{a_sC_A}{6}
\left[(11+2f_A(1-2C^F_A)\right]\left(1-s^{-1}\right),
\nonumber\\
\tilde{P}_{gg}^{(2)}(\omega,a_s) &=& 
C_A a_s \omega \left[K^{gg}_1 (s-1)
-K^{gg}_2 \left(1-s^{-1}\right) -  
K^{gg}_3 \left(1-s^{-3}\right)\right],
\label{qq+gg0+1+2}
\end{eqnarray}
where
$C_A^F = C_F/C_A$, $f_A = 2 n_f T_R/C_A$, and 
\begin{eqnarray}
K^{qq}&=&\frac{1}{18} 
\left\{ \left[51-12f_A\left(7-18C^F_A\right)\right] \frac{L}{s}
 - \left[11-2f_A\left(3-10C^F_A\right)\right] 
\left(1- \frac{s-1}{2\eta} \right)\right.
\nonumber\\
&&{}-\left. \left[51-3f_A\left(1-4C^F_A\right)\right] \frac{s-1}{2}
-20\frac{(s-1)L}{2\eta}
- 2 \left[5-2f_A\left(1-3C^F_A\right)\right] \frac{s-1}{2\eta}
L^2 \right\},
\nonumber\\
K^{gg}_1 &=& \frac{1193}{576}-\zeta_2 - \frac{7f_A}{144} \left(5+2C^F_A\right) 
+\frac{f_A^2}{144} \left[1+4C^F_A\left(1-3C^F_A\right)\right],
\nonumber\\
K^{gg}_2 &=& \frac{415}{288}-\zeta_2 + \frac{f_A}{36} \left(5+2C^F_A\right)
- \frac{f_A^2}{72} \left[1-4C^F_A\left(2-3C^F_A\right)\right],
\nonumber\\
K^{gg}_3 &=& \frac{1}{576} \left[1+2C^F_A\left(1-2C^F_A\right)\right]^2,
\end{eqnarray}
with $\zeta_2=\pi^2/6$.

Through NNLL accuracy, the nondiagonal splitting functions may be represented
as
\begin{eqnarray}
P_{qg}(\omega,a_s)&=&-\frac{\alpha_\omega}{\epsilon_\omega}P_{qq}(\omega,a_s),
\label{qg}\\
P_{gq}(\omega,a_s)&=&-\frac{\epsilon_\omega}{\alpha_\omega}P_{g g}(\omega,a_s)
+ \overline{P}_{gq}^{(1)}(\omega,a_s)
+ \overline{P}_{gq}^{(2)}(\omega,a_s),
\label{gq+1+2}
\end{eqnarray}
where
\begin{eqnarray}
\overline{P}_{gq}^{(1)}(\omega,a_s)&=& 
- \frac{2}{3} C_F a_s  \left[1+f_A\left(1-2C^F_A\right)\right] ,  
\nonumber\\
\overline{P}_{gq}^{(2)}(\omega,a_s) &=& 
C_F a_s \omega \left\{ \frac{1}{9} K^{gq}_1- K^{gq}_2
\left[1-\frac{s-1}{2\eta}(L+2)\right] \right\}, 
\label{qg2}
\end{eqnarray}
with
\begin{eqnarray}
K^{gq}_1 &=& 3 + 5f_A \left(1-2C^F_A\right) - f_A^2 \left(1-2C^F_A\right)^2,
\nonumber\\
K^{gq}_2 &=& 3-\frac{5}{2} C^F_A -4\zeta_2 \left(1-2C^F_A\right)
- \frac{f_A}{18} \left(23-24C^F_A\right)
+ \frac{2f_A^2}{9} C^F_A \left(1-2C^F_A\right).
\end{eqnarray}

We observe from Eq.~(\ref{qg}) that the relation for the NNLL-resummed parts
of the splitting functions $P_{qg}(\omega,a_s)$ and $P_{qq}(\omega,a_s)$ in
Eq.~(\ref{tolja3.1}) is not only correct for their terms singular as
$\omega \to 0$, but also for their regular ones.
The situation is different for the relation between $P_{gq}(\omega,a_s)$ and
$P_{gg}(\omega,a_s)$ in Eq.~(\ref{tolja3}), which does not carry over to the
regular terms, as is evident from Eq.~(\ref{gq+1+2}).
However, the additional terms in Eq.~(\ref{qg2}) have simple forms compared to
the expression for $P_{gg}(\omega,a_s)$ in Eq.~(\ref{qq+gg0+1+2}).

\end{appendix}

\end{document}